\documentclass[letterpaper,titlepage,11pt]{article}
\usepackage{hyperref}
\usepackage{amssymb,amsmath,amsfonts}
\usepackage{epsfig}
\usepackage{graphicx}

\def\clock{{\count0=\time
           \divide\count0 60
           \ifnum\count0<10 0\fi\the\count0
           \multiply\count0 -60 \advance\count0 \time
           :\ifnum\count0<10 0\fi \the\count0
         }}
\newcommand{\timestamp}{{\small\vbox{\hbox{\tt\jobname.tex}
\hbox{\the\day/\the\month/\the\year, \clock}}}}

\newcommand{\be}{\begin{equation}} \newcommand{\ee}{\end{equation}}
\newcommand{\bea}{\begin{eqnarray}} \newcommand{\eea}{\end{eqnarray}}
  \newcommand{\CG}{\mathcal{G}}

\newcommand{\id}{\hbox{1\kern-.27em l}}
\newcommand{\sid}{\hbox{\scriptsize1\kern-.27em l}}
\newcommand{\pa}{\partial} 

\newcommand{\we}{\kern-.1em\wedge\kern-.1em}
\newcommand{\scal}{\kern-.13em\cdot\kern-.13em}

\newcommand{\II}{I\kern-.09em I}

\newcommand{\Z}{\mathbb{Z}} 
\newcommand{\R}{\mathbb{R}}
\newcommand{\T}{\mathbb{T}}

\newcommand{\nn}{\nonumber}

\newcommand{\spa}{\ , \ \ }

\newcommand{\gym}{g_{\mathrm{YM}}}

\newcommand{\vecto}[2]{\left( \begin{array}{c} #1 \\ #2 \end{array}
\right) }
\newcommand{\matrto}[4]{\left( \begin{array}{cc} #1 & #2 \\
#3 & #4 \end{array} \right) }

\newcommand{\matrbig}[4]{\left( \begin{array}{cc} \displaystyle #1 &
\displaystyle #2 \\[3mm]
\displaystyle #3 & \displaystyle #4 \end{array} \right) }

\newcommand{\ds}{\displaystyle}

\numberwithin{equation}{section}

\begin{document}

\begin{titlepage}

\rightline{\vbox{\small\hbox{\tt hep-th/0509011} }}
\vskip 3cm

\centerline{\Large \bf Instabilities of Near-Extremal Smeared
Branes} \vskip 0.1cm \centerline{\Large \bf and the Correlated
Stability Conjecture}

\vskip 1.6cm \centerline{\bf Troels Harmark, Vasilis Niarchos,
Niels A. Obers} \vskip 0.5cm \centerline{\sl The Niels Bohr
Institute} \centerline{\sl Blegdamsvej 17, 2100 Copenhagen \O,
Denmark}

\vskip 0.5cm

\centerline{\small\tt harmark@nbi.dk, niarchos@nbi.dk, obers@nbi.dk}

\vskip 1.6cm

\centerline{\bf Abstract} \vskip 0.2cm \noindent We consider the
classical and local thermodynamic stability of non- and
near-extremal D$p$-branes smeared on a transverse direction. These
two types of stability are connected through the correlated
stability conjecture for which we give a proof in this specific
class of branes. The proof is analogous to that of Reall for
unsmeared branes, and includes the construction of an appropriate
two-parameter off-shell family of smeared D$p$-brane backgrounds.
We use the boost/U-duality map from neutral black strings to
smeared black branes to explicitly demonstrate that non-and
near-extremal smeared branes are classically unstable, confirming
the validity of the conjecture. For near-extremal smeared branes
in particular, we show that a natural definition of the grand
canonical ensemble exists in which these branes are
thermodynamically unstable, in accord with the conjecture.
Moreover, we examine the connection between the unstable
Gregory-Laflamme mode of charged branes and the marginal modes of
extremal branes. Some features of T-duality and implications for
the finite temperature dual gauge theories are also discussed.

\vskip 0.5cm

\end{titlepage}

\pagestyle{empty}
\small
\tableofcontents
\normalsize

\pagestyle{plain}
\setcounter{page}{1}

\section{Introduction}

The gauge/gravity correspondence \cite{Aharony:1999ti} implies a
deep connection between the thermodynamics of near-horizon brane
backgrounds and that of the dual non-gravitational theories. In
particular, this suggests that the gauge theory dual of a
classically unstable brane background must have a corresponding
phase transition. This hints at an interesting connection between
the classical stability of brane backgrounds and the thermodynamic
stability of the dual non-gravitational theories, and thereby to
the thermodynamic stability of the brane backgrounds themselves.

Gregory and Laflamme \cite{Gregory:1993vy,Gregory:1994bj} were the
first to study the classical stability of neutral black strings,
as well as certain charged branes. They showed that neutral black
strings are unstable under perturbations that oscillate in the
direction in which the string extends, provided that the
wavelength of the perturbation is larger than the order of the
horizon radius. When the extended direction is compactified on a
circle, this means in particular that for masses below/above a
critical mass the black string is unstable/stable. Already in this
original work, a global thermodynamic argument for the instability
was given, namely that for small masses the entropy of a localized
black hole is higher than that of the black string with the same
mass, suggesting a possible endpoint
for the decay of the black string in that case.%
\footnote{The viewpoint that an unstable uniform black string decays to
a localized black hole has been challenged in
\cite{Horowitz:2001cz} (in this connection see also the discussion
in the reviews \cite{Harmark:2005pp,Kol:2004ww}).}

This deep connection between classical and thermodynamic stability
of brane solutions was recently made more precise with a
conjecture formulated by Gubser and Mitra
\cite{Gubser:2000ec,Gubser:2000mm}. The conjecture, referred to as
the Correlated Stability Conjecture (CSC), states that for systems
with translational symmetry and infinite extent, a
Gregory-Laflamme (GL) instability arises precisely when the system
has a local thermodynamic instability. In further detail, local
thermodynamic stability is defined here as positive-definiteness
of the Hessian matrix of second derivatives of the mass with
respect to the entropy and any charges that can be redistributed
over the direction in which the GL instability is supposed to
occur. Thus, the CSC relates a local thermodynamic instability to
a perturbative dynamical instability. The latter is the classical
gravitational instability arising from perturbations of the brane
background.

For magnetically charged D$p$-brane solutions in String/M-theory,
a semi-classical proof of the CSC was given by Reall \cite{Reall:2001ag}
using the Euclidean path integral formulation of gravity.%
\footnote{See also \cite{Gregory:2001bd}. See furthermore
\cite{Gregory:1994tw,Hirayama:2002hn,Kang:2004hm,Kang:2004ys} for
work on the classical stability of charged branes.} A key
ingredient in this proof is the relation between the threshold
mode of the classical instability and a Euclidean negative mode in
the semi-classical path integral. In this way, for this class of
branes it was shown that a classical instability appears precisely
when the specific heat becomes negative. The CSC has also been
considered and confirmed recently in more complicated settings
involving bound states of branes
\cite{Gubser:2004dr,Ross:2005vh,Friess:2005tz}. A class of
counter-examples in a different setting appeared very recently in
\cite{Friess:2005zp}. Further comments related to these examples
can be found in Section \ref{sec:csc}.

Another interesting class of branes is that of smeared
D$p$-branes. Here, we call an electrically charged brane smeared
if it has a direction with translational symmetry along which the
brane is not charged. For magnetically charged branes we follow
the opposite convention. When compactified on the isometric
direction, a smeared D$p$-brane is related by T-duality to a
D$(p+1)$-brane wrapping the T-dual circle. In particular, the
near-extremal limit of the latter has a dual description in terms
of a $(p+1)$-dimensional supersymmetric Yang-Mills (SYM) theory
compactified on a circle. Thus, the issue of stability of smeared
black branes has non-trivial implications for the stability of the
dual SYM theories on $\R^{p-2} \times \T^2$, where the torus
consists of the compact spatial world volume direction and the
Euclidean time direction.

Moreover, there is an interesting connection between neutral black
strings and smeared D$p$-branes. Building on the original
observation of Ref.~\cite{Harmark:2002tr}, it was shown in
\cite{Harmark:2004ws} (see also
\cite{Bostock:2004mg,Aharony:2004ig,Kudoh:2005hf}) that the phases
of Kaluza-Klein black holes (see the reviews
\cite{Harmark:2005pp,Kol:2004ww,Harmark:2005pq}) can be mapped
onto phases of non- and near-extremal D$p$-branes with a circle in
their transverse space. The map is a sequence of boost and
U-duality transformations, and includes as a special case the map
of neutral black strings to non- and near-extremal D$p$-branes
that are uniformly smeared on a transverse circle.

For neutral black strings wrapped on a circle
a new static phase of non-uniform strings
\cite{Gregory:1988nb,Gubser:2001ac,Wiseman:2002zc,Sorkin:2004qq}
is known to emerge at the GL transition
point, where the unstable mode is at threshold.
In Refs.~\cite{Bostock:2004mg,Aharony:2004ig,Harmark:2004ws}
it was observed that the GL point is mapped onto a
critical mass/energy for non-/near-extremal branes, where a new
phase of non- and near-extremal branes, non-uniformly distributed
on the circle, emerges. This picture strongly suggests that
 non- and near-extremal smeared branes exhibit classical
instabilities.  In particular, the threshold mode of the black
string is mapped directly onto the threshold mode of  non-
and near-extremal smeared black D$p$-branes. More generally,
it is natural to expect that the (time-dependent) unstable mode
can also be obtained from the neutral GL unstable mode by a suitable
generalization of the boost/U-duality map \cite{Aharony:2004ig}.

The aim of this paper is to study the classical stability
and CSC for  non- and near-extremal smeared branes in detail.%
\footnote{Our brane solutions do not include Kaluza-Klein bubbles.
The classical stability of another class of smeared branes,
involving bubbles, has been considered in
Ref.~\cite{Sarbach:2004rm}. For solutions involving branes and
bubbles, see also \cite{Harmark:2004bb}.} Our analysis will show
in a quantitative way how the presence of charge affects the GL
instability of neutral black strings, and what happens when we
take the near-extremal limit. An important tool that we use is the
boost/U-duality map mentioned above. In this work we restrict
ourselves to smeared  D$p$-branes of type II string theory, but
the other 1/2-BPS branes can be treated similarly.

The main points of this paper can be summarized as follows:
\begin{itemize}
\item We argue that when applying the CSC to smeared branes one
should consider thermodynamic stability in the grand canonical
ensemble since the charge can redistribute itself in the direction
in which the brane is smeared. Non-extremal smeared branes are
thermodynamically unstable in this ensemble and according to the
CSC these branes should have a classical instability. The choice
of ensemble is independent of whether the brane is electrically or
magnetically charged.
\item We give a proof of the CSC for smeared
branes. One of the essential elements is the explicit construction
of an appropriate off-shell two-parameter family of Euclidean
backgrounds connected to these branes.
\item Following Ref.~\cite{Aharony:2004ig}, we give an explicit
construction of the GL unstable mode for non-extremal smeared
branes. More specifically, we map the time-dependent unstable GL
mode of neutral black strings to a time-dependent unstable mode of
non-extremal smeared branes. This confirms the validity of the CSC
for non-extremal smeared branes.
\item We present a detailed analysis of the near-extremal limit
and the validity of the CSC in this case. We show explicitly that
the near-extremal limit of the unstable GL mode of non-extremal
smeared branes is well-defined and hence that near-extremal
smeared  branes are classically unstable. According to the CSC
this means that near-extremal smeared branes are thermodynamically
unstable in the grand canonical ensemble. Nevertheless, it seems
that in the near-extremal limit the charge cannot vary anymore,
and it is not a priori obvious how to define a grand canonical
ensemble for these branes. We show, however, that a natural
definition of such an ensemble exists. This implies a new version
of the first law of thermodynamics for near-extremal smeared
branes, involving a rescaled chemical potential. In this
near-extremal grand canonical ensemble, the near-extremal smeared
branes are thermodynamically unstable. In this way, we resolve an
apparent puzzle, showing that the CSC is also
correct for near-extremal branes.%
\footnote{Note that this also resolves the contradiction found in
the numerical investigations of Ref.~\cite{Kang:2005is}.}
\item We examine the connection between the GL mode for charged
branes and marginal modes for extremal smeared branes. We show
that in the extremal limit the GL mode of near-extremal smeared
branes precisely becomes the marginal mode of extremal smeared
branes. As a consequence of this fact the extremal smeared branes
are in a sense arbitrarily close to being unstable.
\item We discuss various issues related to T-duality on smeared
D$p$-branes and the properties of the dual gauge theories.
\end{itemize}

The outline of this paper is as follows. We start in
Section \ref{sec:prem} by reviewing how non- and
near-extremal smeared D$p$-branes of type II string theory
can be obtained from the neutral black string solutions using the boost/U-duality
map. We also review the main results on the thermodynamics of these branes,
which we will need when considering the CSC.

In Section \ref{sec:csc}, we will review the CSC and discuss the
criteria that determine the choice of thermodynamic
ensemble. Then following the arguments of Reall \cite{Reall:2001ag},
we present a proof of the CSC for smeared
branes. Some of the subtle issues that are not fully resolved will be
discussed as well. We also comment on the general proof of the CSC
and the validity of the CSC in the presence of compact directions.

We continue in Section \ref{sec:mode} by showing that, with a
suitable generalization of the boost/U-duality transformation
reviewed in Section \ref{sec:prem}, we can map the GL unstable mode of the
neutral black string to a time-dependent unstable mode for
non-extremal smeared branes. This map relies on an argument first noticed
in Ref.~\cite{Aharony:2004ig}.
We then take the near-extremal limit of this and show that it gives a
well-defined unstable mode for near-extremal smeared branes. As a
consequence we conclude that the latter are also classically
unstable.

In Section \ref{sec:ext} we consider the extremal case. We show
that the extremal branes have marginal modes of any wave-length
and that these modes can be recovered by taking the extremal limit
of the GL mode of charged smeared branes. This means that when an
extremal smeared brane is perturbed in such a way that it becomes
non-BPS the resulting brane background becomes unstable.
Furthermore, for any marginal perturbation of the extremal smeared
branes one can find a corresponding non-BPS continuation.

We return to the CSC in Section \ref{sec:puz},
where we give the appropriate definition of the grand canonical ensemble
in the near-extremal case. We then show that, in this ensemble,
near-extremal branes are thermodynamically unstable. Together with
the explicit construction of the unstable mode
for near-extremal branes in Section \ref{sec:mode},
this confirms the validity of the CSC in this case as well.

We conclude with a short summary and some open questions
in Section \ref{sec:con}. Two appendices are also included. In Appendix
\ref{app:mode} we present the differential equations determining the neutral GL mode. In
Appendix \ref{app:fam} we provide the details of the off-shell
two-parameter family of black brane backgrounds that is essential
for the proof  of the CSC for smeared branes in Section \ref{sec:csc}.

\section{Preliminaries}
\label{sec:prem}

In this section we review how the non- and near-extremal smeared
D$p$-branes can be obtained from the neutral black string by using
the boost/U-duality map, and the near-extremal limit. For later
use throughout the paper, we also review the main results on the
thermodynamics of these branes.

\subsection{Non-and near extremal smeared branes from boost/U-duality
\label{sec:boost}}

We begin by recalling how one obtains the non-extremal smeared D$p$-brane
solution of type II string theory from the uniform black string
solution in pure gravity, by applying a sequence of transformations
including a boost and U-dualities.%
\footnote{This method of ``charging up'' neutral solutions
was originally conceived in \cite{Hassan:1992mq} where it
was used to obtain black $p$-branes from neutral black
holes.}

The starting point is the metric of the uniform black string in $10-p$ dimensions:
\begin{equation}
\label{bsmet} ds^2 = - f dt^2 + dz^2 + f^{-1} dr^2 + r^2
d\Omega_{7-p}^2 \spa f = 1 - \frac{r_0^{6-p}}{r^{6-p}} \ .
\end{equation}
By adding $p+1$ flat directions, this solution is trivially uplifted
to a solution of eleven-dimensional (super)gravity
\begin{equation}
\label{bsmet11} ds_{11}^2 = - f dt^2 + dz^2 + f^{-1} dr^2 + r^2
d\Omega_{7-p}^2 + \sum_{i=1}^{p} dx_i^2 + dy^2 \ .
\end{equation}
Here, the coordinates $x_i$ are used for $p$ flat directions and
$y$ for the one that will be taken to be the eleventh direction
when reducing from M-theory to type IIA string theory.
The metric \eqref{bsmet11} is thus a vacuum solution of M-theory.
Performing a Lorentz-boost in the $y$-direction
\begin{equation}
\label{boost} \vecto{t_{\rm new}}{y_{\rm new}} = \matrto{\cosh
\alpha}{\sinh \alpha}{\sinh \alpha}{\cosh \alpha} \vecto{t_{\rm
old}}{y_{\rm old}} \spa \alpha >0  \ ,
\end{equation}
and dropping the label `new' gives the boosted metric
\begin{equation}
\label{boomet}
\begin{array}{rcl} \ds
ds^2_{11} &=& \ds \left( - f \cosh^2 \alpha + \sinh^2 \alpha
\right) dt^2 + 2 (1-f) \cosh \alpha \sinh \alpha \, dt dy
\\[1mm]
&& \ds + \left( - f \sinh^2 \alpha + \cosh^2 \alpha \right) dy^2 +
dz^2 + f^{-1} dr^2 + r^2 d\Omega_{7-p}^2  + \sum_{i=1}^{p}
(dx_i)^2 \ .
\end{array}
\end{equation}
Since we have an isometry in the $y$-direction we can now make an
S-duality in the $y$-direction to obtain a solution of type IIA
string theory. This gives a non-extremal D0-brane solution of type
IIA string theory which is uniformly smeared along the $\R^{p+1}$
space parameterized by $x_i$ and $z$. With a T-duality
transformation on each of the $p$ directions $x_i$, we deduce the
solution for a non-extremal D$p$-brane smeared along the
$z$-direction,
\begin{equation}
\label{metD0a}
\begin{array}{c}
\ds ds^2 = H^{-1/2} \left( - f dt^2 + \sum_{i=1}^p dx_i^2 \right)
+ H^{1/2} \left( f^{-1} dr^2 + dz^2  + r^2 d\Omega_{7-p}^2 \right)
\ ,
\\[5mm] \ds
H = 1 +  \frac{r_0^{6-p}\sinh^2 \alpha}{r^{6-p}} \spa e^{2\phi} =
H^{\frac{3-p}{2}}  \spa  A_{01\ldots p} =  \coth \alpha \, (
H^{-1} - 1 ) \ ,
\end{array}
\end{equation}
written in the string frame. The function $f(r)$ is given in
\eqref{bsmet}. One can also use further U-dualities to obtain the
backgrounds for smeared F1-strings and NS5-branes, but we choose
to focus on D-branes in this paper.

By one further T-duality in the transverse $z$-direction, we can relate the
D$p$-brane smeared on a circle to that of a D$(p+1)$-brane wrapped on
the T-dual circle. Since
this T-duality is important for defining the near-extremal limit
and computing quantities in the dual gauge theories of near-extremal
D$p$-branes smeared on a transverse circle, we give the relevant T-duality relations here.
Assuming that the coordinate $z$ in \eqref{metD0a} is compactified on a circle
of circumference $L$, $i.e.$ $z \sim z + L$, and denoting the string coupling by
$g_s$, we find that the circumference $\bar L$ of the T-dual circle
(on which the D$(p+1)$-brane is wrapped) and the T-dual string coupling
$\bar g_s$ are related by
\begin{equation}
L \bar{L} = (2\pi l_s)^2 \spa g_s = \bar{g}_s \frac{2\pi
l_s}{\bar{L}} \ ,
\end{equation}
where $l_s$ is the string length.
Introducing the gauge theory quantities $g_{\rm YM}^2 = (2\pi)^{p-1}
\bar g_s l_s^{p-2}$ and $\lambda = g_{YM}^2 N $ for the $(p+1)$-dimensional
supersymmetric Yang-Mills compactified on a spatial circle, it also follows
that we have the relation
\begin{equation}
\label{Kdef} r_0^{6-p} \cosh \alpha \sinh \alpha = K l_s^{2(4-p)}
\spa K \equiv \frac{\lambda (2\pi)^{7-2p}}{(6-p)\Omega_{7-p}} \ .
\end{equation}
Another useful definition is the following
\begin{equation}
\label{Gdef} \frac{L V_{p} \Omega_{7-p} }{16 \pi G} =
\frac{1}{{\cal{G}}} l_s^{-2(6-p)} \spa \frac{1}{{\cal{G}}} \equiv
(2\pi)^{2p-9} \bar{L} V_p \Omega_{7-p} \frac{N^2}{\lambda^2} \ ,
\end{equation}
where we used the T-duality and gauge theory relations given above along with
the definition $16\pi G = (2\pi)^7 g_s^2 l_s^8$.

\subsubsection*{Near-extremal limit}

We define the near-extremal limit as
\begin{equation}
\label{nelim} l_s \rightarrow 0 \spa u = \frac{r}{l_s^2} \spa
\hat{z} = \frac{z}{l_s^2} \spa g_{\rm YM}, \ \bar{L} \
\mbox{fixed} \ ,
\end{equation}
where $(r,z)$ are the coordinates appearing in the non-extremal
background \eqref{metD0a}. Using this limit, the resulting
solution for near-extremal
 smeared D$p$-branes is
\begin{equation}
\label{NEbrane}
\begin{array}{c}
\ds l_s^{-2} ds^2 = \hat{H}^{-1/2} ( - f dt^2 + \sum_{i=1}^p
dx_i^2 ) + \hat{H}^{1/2} ( f^{-1} du^2 + d\hat{z}^2 + u^2
d\Omega_{7-p}^2 ) \ ,
\\[3mm] \ds
e^{2\phi} = \hat H^{\frac{3-p}{2}}  \spa A_{01\ldots p} = \hat
H^{-1} \spa \hat{H} = \frac{K}{u^{6-p}} \spa f = 1 -
\frac{u_0^{6-p}}{u^{6-p}} \ ,
\end{array}
\end{equation}
where $K$ is defined in \eqref{Kdef}.

The classical and local thermodynamic stability of the non- and
near-extremal D$p$-brane backgrounds given above is the main
object of our investigation in this paper. The boost/U-duality
transformation and the near-extremal limit described above will
prove very useful in this context.

We recall here that the sequence of boost and U-duality transformations reviewed above
can be equally well applied to other solutions of pure gravity with
a compact circle direction. In particular, in Ref.~\cite{Harmark:2004ws}
this map is discussed in great detail, and applied to both
black holes on cylinders and non-uniform
black strings, generating non- and near-extremal black branes that
are either localized on a transverse circle or non-uniformly distributed on
the circle.

\subsection{Thermodynamics of non- and near-extremal smeared branes
\label{sec:thermo}}

We now review some important results on the
thermodynamics of the non- and
near-extremal smeared D$p$-branes obtained above.

The thermodynamics of the non-extremal D$p$-branes \eqref{metD0a}
smeared on a transverse circle of circumference $L$, is given by
\begin{equation}
\label{thermon1}
\begin{array}{c}
\ds \frac{M}{L} =\frac{\Omega_{7-p}}{16 \pi G} V_p
r_0^{6-p}[7-p+(6-p)\sinh^2 \alpha] \spa
 \frac{S}{L}=4\pi \frac{\Omega_{7-p} }{16
\pi G} V_p r_0^{7-p} \cosh \alpha \ ,
\\[4mm] \ds
 \frac{Q}{L}=\frac{\Omega_{7-p} }{16 \pi G} V_p r_0^{6-p} (6-p)
\sinh \alpha \cosh \alpha \spa
 T=\frac{6-p}{4\pi r_0 \cosh \alpha} \spa
 \nu=\tanh \alpha  \ ,
\end{array}
\end{equation}
where $V_p$ is the world-volume of the brane.
Here the extensive quantities, $M/L$, $Q/L$ and $S/L$ correspond
respectively to the mass density, charge density and entropy density
along the transverse $z$ direction. $T$ is the temperature and $\nu$ the
chemical potential. These quantities satisfy the usual first law of
thermodynamics $dM = S dT + \nu d Q$.

Correspondingly, the thermodynamic quantities of near-extremal
D$p$-branes \eqref{NEbrane} are given by
\begin{equation}
\label{thermon3}
E =  \frac{1}{{\cal{G}}} u_0^{6-p} \frac{8-p}{2} \spa
 S=  \frac{4\pi}{{\cal{G}}} u_0 ( u_0^{6-p} K)^{1/2} \spa T  =
\frac{6-p}{4\pi u_0} \left( \frac{u_0^{6-p}}{K} \right)^{1/2} \ .
\end{equation}
Here, $E$ is the energy above extremality defined by $E = \lim (M- Q)$
in the near-extremal limit \eqref{nelim}, while $K$ and ${\cal{G}}$,
defined in \eqref{Kdef}, \eqref{Gdef} are finite in the limit.
The quantities in \eqref{thermon3} satisfy the usual first law of
thermodynamics $dE = S dT$.

\subsubsection*{Canonical and grand canonical ensemble}

In what follows, we will make extensive use of
both the canonical as well as grand canonical ensemble, so we briefly
review the definition of these ensembles here.

In the canonical ensemble, the temperature and charge are kept fixed and the
appropriate thermodynamic potential is the Helmholtz free
energy
\begin{equation}
F (T,Q) = M - T S \spa d F = -S d T  + \nu d Q \ .
\end{equation}
The condition for thermodynamic stability in the canonical
ensemble is
\begin{equation}
\label{specheat} C_{Q} \equiv \left( \frac{\partial M}{\partial T}
\right)_Q = T \left( \frac{\partial S}{\partial T} \right)_Q > 0 \
,
\end{equation}
$i.e.$ positivity of the specific heat $C_Q$.

In the grand canonical ensemble, the temperature and chemical potential are kept
fixed, so the appropriate thermodynamic potential is the Gibbs free energy
\begin{equation}
G (T,\nu)= M - TS - \nu Q \spa dG = - S dT - Q d\nu \ .
\end{equation}
The condition for thermodynamic stability in the grand canonical ensemble
is
\begin{equation}
\label{grand}
C_Q \equiv \left( \frac{\partial M}{\partial T} \right)_Q
 > 0
\spa c \equiv \left( \frac{\partial \nu}{\partial Q} \right)_T > 0
\ ,
\end{equation}
where $C_Q$ is the specific heat and $c$ is the inverse isothermal electric
permittivity. The condition \eqref{grand} follows from demanding that the
Hessian $H_G$ of the Gibbs free energy
is negative definite. In fact, because of the matrix relation $H_G= - H_M^{-1}$,
where $H_M$ is the Hessian of the mass $M(S,Q)$, this
is equivalent to demanding that $H_M$ is positive
definite, as required by local thermodynamic equilibrium.

Using the thermodynamic quantities for non-extremal branes in
\eqref{thermon1} and the definitions in \eqref{grand} one computes in this case%
\footnote{These expressions are most easily computed using the
identity that $\frac{ \partial y(r,s)}{\partial x(r,s)} \vert_{z(r,s)}
= \frac{ \partial(y,z)} {\partial(r,s)}
[\frac{ \partial(x,z)}{\partial (r,s)}]^{-1}.
$}
\begin{equation}
\label{cqnex}
C_Q= \left[ \frac{ 7-p  + (8-p) \sinh^2 \alpha}{-1 + (4-p) \sinh^2
\alpha} \right] S \spa c =   \frac{1}{ \cosh^2 \alpha [1 -(4-p)
\sinh^2 \alpha]}\frac{1}{TS} \ .
\end{equation}
With these results and the conditions for thermodynamic stability
reviewed above, it is not difficult to determine the conditions
for thermodynamic stability of non-extremal smeared D$p$-branes.
For $p \leq 3$, positivity of the specific heat becomes
\begin{equation}
\label{cq2}
C_Q > 0 \qquad \Leftrightarrow \qquad \alpha
> {\rm arcsinh}  (1/\sqrt{4-p}) \ ,
\end{equation}
showing that there is a lower bound on the charge of the branes in order
to be thermodynamically stable in the canonical ensemble.
On the other hand, we have
\begin{equation}
\label{cq3} c > 0 \qquad \Leftrightarrow \qquad 0 < \alpha <  {\rm
arcsinh}  (1/\sqrt{4-p}) \ ,
\end{equation}
which is incommensurate with the condition in \eqref{cq2}. For
$p=4,5$ we have that $C_Q < 0$ and $c>0$. Hence for any
$p=0,1,...,5$ it is impossible to satisfy the requirement
\eqref{grand} of thermodynamic stability in the grand canonical
ensemble.

We now turn to the case of near-extremal branes.
Since the charge has been sent to infinity, there seems to
be for this case only one relevant ensemble,
namely the canonical ensemble. The corresponding Helmholtz potential
is given by
\begin{equation}
F (T) = E - TS  \spa d F = -S d T \ ,
\end{equation}
and thermodynamic stability requires positivity of the specific heat
$C \equiv \partial E(T)/\partial T$.
 From the thermodynamics \eqref{thermon3} one
easily obtains the simple relation
\begin{equation}
\label{cqneex} C = \frac{8-p}{4-p} S \ ,
\end{equation}
As a check, the same result is obtained by taking the near-extremal limit
($\alpha \rightarrow \infty$) of the expression \eqref{cqnex} for $C_Q$.
For any non-zero temperature, the specific heat is thus positive for
near-extremal smeared D$p$-branes when $p \leq 3$, which is a well-known result.

Since in the near-extremal limit the charge goes to infinity and
the chemical potential goes to one, it seems that these quantities
cannot be varied anymore. This suggests that it does not make
sense to consider the grand canonical ensemble for near-extremal
branes. Moreover, one could consider what happens to the quantity
$c$ in \eqref{cqnex} when taking the near-extremal limit. Since
$\alpha \rightarrow \infty $ in the near-extremal limit, we
clearly have that $c \rightarrow 0$ in the limit. This would seem
to imply that, infinitesimally close to the near-extremal limit,
non-extremal branes are marginally thermodynamically stable in the
grand canonical ensemble. As we will see below when we consider
the correlated stability conjecture, which relates classical and
thermodynamic stability, this fact implies a puzzle in view of the
fact that near-extremal branes are classically unstable. However,
in Section \ref{sec:puz} we will demonstrate that there is a
natural resolution of this apparent violation of the CSC.

\subsubsection*{T-duality}

Finally we have a few remarks on the effect of the T-duality along
the $z$-direction (see Section \ref{sec:boost}) on the
thermodynamics. Under this transformation the thermodynamic
quantities \eqref{thermon1} and \eqref{thermon3} for non- and
near-extremal branes are invariant. Thus, the thermodynamics of a
D$p$-brane smeared on a circle is identical to that of a
D$(p+1)$-brane wrapped on the T-dual circle. Despite this fact, in
the next section we will see that when applying the CSC there is a
qualitative distinction between the two cases.


\section{The correlated stability conjecture for charged branes
\label{sec:csc}}

\subsection{Formulation of the conjecture \label{sec:conj}}

In \cite{Gubser:2000ec,Gubser:2000mm} Gubser and Mitra put forward
an intriguing conjecture that relates classical and thermodynamic
instabilities in gravitational systems. The precise form of this
conjecture states that a gravitational system with a non-compact
symmetry ($e.g.$ a black brane with non-compact worldvolume) is
classically stable, if and only if, it is locally
thermodynamically stable. For a system with $n$ conserved charges
$Q_i$ this criterion of local thermodynamic stability translates
(after the appropriate choice of ensemble) into a positivity
criterion for the $(n+1) \times (n+1)$ Hessian matrix
$H_M=\big(\frac{\pa ^2 M}{\pa q_\alpha \pa q_\beta}\big)$, which
involves the second derivatives of the mass $M$ with respect to
the entropy $S$ and the charges $Q_i$. A partial proof of this
conjecture has been given by Reall in \cite{Reall:2001ag}, who
considered the case of magnetically charged black branes. However,
a general proof of the conjecture is yet to be found.

Before even applying the CSC one should first make
the appropriate choice of thermodynamic ensemble.
This choice is important, because it affects crucially the discussion of local thermodynamic
stability and ultimately the validity of the conjecture as such.
The usual practice is to consider gravity in the canonical
ensemble, where the temperature $T$ is kept fixed
and the partition function of the system is expressed
as a function of $T$. In general situations
with an arbitrary number of charges $Q_i$, however,
the choice of the ensemble is not always obvious
and part of the conjecture should involve a clear statement
that singles out the right choice.
In what follows, we attempt to illuminate this point for
a large class of smeared or unsmeared, magnetically or
electrically charged black branes.

To be more concrete, consider a generic (non-extremal)
black D$p$-brane solution in type II string theory.
By $\{z_i\}$ let us denote the set
of non-compact directions, along which the brane
exhibits translational symmetry.
This solution may be charged electrically or magnetically
under a $(p+1)$-form gauge field, with a corresponding charge $Q_p$.
An electrically charged brane will be called smeared along
$z_i$, whenever it is not charged along this direction.
For magnetically charged branes, we follow the opposite
convention and call the brane smeared along $z_i$,
whenever it is charged along $z_i$.\footnote{{This
definition is also very natural for
D3-branes which are self-dual under electric-magnetic duality.}}
In principle, the choice of thermodynamic ensemble depends
crucially on whether we treat $Q_p$ as a parameter
that has been fixed or as a parameter
that can vary freely.

In the first case, we consider the system in the {\it canonical}
ensemble, where the partition function depends on the temperature
and the fixed charge. As explained in Section \ref{sec:thermo}, in
that case local thermodynamic stability requires the positivity of
the specific heat (\ref{specheat}). The situation is slightly
different, when the charge is allowed to vary freely. Then, we
keep the corresponding chemical potential $\nu$ fixed and discuss
local thermodynamic stability in the {\it grand canonical}
ensemble. In this ensemble the partition function depends on $\nu$
and the temperature $T$ and the Hessian of the mass is positive
definite precisely when (\ref{grand}) is satisfied.

The proposal is that thermodynamic computations should be
done in the grand canonical ensemble with respect to the charge $Q$, when
the branes are smeared along at least one of the directions $\{z_i\}$.
The opposite should be true when the brane is not smeared
in any of the directions $\{z_i\}$. Note that this statement
applies equally well to electrically or magnetically
charged branes. Indeed, any sensible formulation of the
CSC should be invariant under the electric-magnetic
duality.

As a simple illustration, consider a non-extremal
D0-brane solution smeared on a
transverse direction $z$ in the type IIA theory. The corresponding
supergravity solution in the string frame takes the form
\begin{equation}
\begin{array}{c}
\displaystyle \label{aac} ds^2=-H^{-1/2} f dt^2+H^{1/2}\bigg(
f^{-1} dr^2 +r^2 d\Omega_7^2 + dz^2\bigg) \spa e^{2\phi}=H^{3/2}\
,
\\[2mm]
\displaystyle  A_0=\coth \alpha (H^{-1}-1) ~,
f(r)=1-\frac{r_0^6}{r^6}~, ~ H(r)=1+\sinh^2 \alpha (1-f)\ .
\end{array}
\end{equation}
The D0-brane charge $Q$ is smeared along the
$z$-direction in this example and as a result
it can be redistributed there freely. Hence,
it is appropriate to consider local thermodynamic stability in
the grand canonical ensemble, where $Q$ is allowed to
vary.

With a T-duality transformation along $z$
the background (\ref{aac}) turns into the D1-brane solution
\begin{eqnarray}
\label{aad} ds^2&=&H^{-1/2} \bigg(-f dt^2+dz^2\bigg)+H^{1/2}\bigg(
f^{-1} dr^2 +r^2 d\Omega_7^2\bigg)~,
\\ \nonumber
e^{2\phi}&=&H~, ~ ~ A_{0z}=\coth \alpha (H^{-1}-1) ~.
\end{eqnarray}
The D1-brane is now charged along the $z$-direction and the
corresponding charge is a fixed quantity. Accordingly, we should
now consider local thermodynamic stability in the canonical ensemble.

As another example consider the D0-D2 bound state
\cite{Gubser:2004dr,Ross:2005vh,Friess:2005tz}. In this system,
the D0-brane is fully embedded inside the D2-brane and smeared
along its two transverse directions inside the worldvolume of the D2-brane.
Then, according to the previous discussion, the CSC should be applied
in the grandcanonical ensemble with respect to the charge $Q_0$
of the D0-brane, but in the canonical ensemble with respect to
the charge $Q_2$ of the D2-brane.

The above choice of thermodynamics fits very nicely with the
classical stability properties of these solutions. As we see
explicitly in later sections, the smeared D0-branes exhibit a GL
instability, which persists even in the near-extremal limit. The
thermodynamic instability arises naturally in the grand canonical
ensemble, but is absent in the canonical ensemble above some
critical value of the charge where the specific heat is strictly
positive (see eq.\ \eqref{cq2}). The D1-branes, on the other hand,
are known to be stable in supergravity for large enough charges.
This feature is captured correctly by the local thermodynamic
stability analysis in the canonical ensemble.

More generally, it is well-known in thermodynamics
that one can move back and forth between the
canonical and grand canonical ensembles with the appropriate
Legendre transform. Here we see that within the context of the
CSC it is natural to associate a Legendre transform with
a T-duality transformation in supergravity.
The appearance of the Legendre transform is
an essential feature of supergravity that
treats momentum and winding modes asymmetrically.
In the full string theory, where momentum and winding
are exchanged by T-duality the Legendre transform
would be unnecessary. A momentum instability for
a smeared brane would transform under T-duality into
a winding instability for a wrapping brane.%
\footnote{Similar statements on this point appeared in
\cite{Ross:2005vh}.}

\subsection{The CSC for smeared branes}

It is interesting to consider the proof of the CSC for smeared
black branes in the grand canonical ensemble. This is a first step
towards the extension of the proof of \cite{Reall:2001ag} in more
general situations, where besides the mass one can also vary an
additional set of charges.

More specifically, consider the case of a non-extremal smeared
D$p$-brane on a transverse direction $z$,  given in
Eqs.~\eqref{metD0a}. In order to address the relation between
classical stability and thermodynamics we repeat the basic
elements of the argument that appears in \cite{Reall:2001ag} (we
refer the reader to that paper for a more detailed description).
In a nutshell, one has to show the following facts (see also the
recent discussion in \cite{Friess:2005tz}): \noindent
\begin{enumerate}
\item[($a$)] Demonstrate the existence of an
appropriate family of Euclidean backgrounds for which the
Euclidean action (relative to flat space) takes the form
\begin{equation}
\label{aaf} I(x,y;\beta,\nu)=\beta(E(x,y)-\nu Q(x,y))-S(x,y) ~.
\end{equation}
The generic point in this family is an off-shell background, which
does not satisfy the Einstein equations of motion, but satisfies
the appropriate Hamiltonian constraints. In the present case, we
consider a two-parameter family of backgrounds parameterized by
two variables, $x$ and $y$. This should be contrasted to the
unsmeared case analyzed in \cite{Reall:2001ag}, where the
corresponding family is one-dimensional. In (\ref{aaf}) $\beta$ is
the inverse temperature and $\nu$ the chemical potential. For some
value of $(x,y)=(x(T,\nu),y(T,\nu))$ the background becomes an
exact solution of the equations of motion. At this point the
parameters $E$, $S$ and $Q$ are precisely the energy, entropy and
charge of the corresponding black brane. For other values of
$(x,y)$ the interpretation of the functions $E$, $S$ and $Q$ is
not important. An explicit construction of this two-parameter
family will be discussed in a moment. \item[($b$)] Verify that the
norm of the on-shell perturbations on the space of theories is
positive. This ensures that the analysis is restricted to
normalizable on-shell perturbations excluding any non-physical
negative-norm conformal perturbations. The precise form of the
norm is related to the Lichnerowicz operator and can be defined as
follows. For compactness, let us denote the field perturbations
collectively by $\Psi^I$. $I$ is in general a multi-index label
and the fields $\Psi^I$ may include scalar, metric and gauge field
perturbations. The ansatz for a static GL mode is $\Psi^I={\rm
Re}(\psi^I e^{ikz})$ and the norm $||\psi^I||^2=\int d^d x \psi^I
\CG_{IJ} \psi^J$ on the space of perturbations is defined through
the metric $\CG_{IJ}$. This metric appears in the linearized
equations of motion in the following way
\begin{equation}
\label{aao} \int d^d x' \frac{\delta^2 I}{\delta \psi^I(x)\delta
\psi^J(x')} \psi^J(x')=-k^2 \CG_{IJ}\psi^J ~.
\end{equation}
One must check that $||\psi^I||^2$ is positive definite.
This will ensure that the action $I$ decreases, and therefore
an instability exists, precisely when the Hessian
$\frac{\delta^2 I}{\delta \psi^I \delta \psi^J}$
fails to be positive definite. Indeed, for quadratic perturbations
\begin{equation}
\Delta I=\frac{1}{2}\int d^d x d^d x' \psi^I(x)
\frac{\delta^2 I}{\delta \psi^I(x)\delta \psi^J(x')} \psi^J(x')
+{\cal O}(\psi^3)=-\frac{k^2}{2}||\psi||^2+{\cal O}(\psi^3)
~.
\end{equation}
\item[($c$)] A final point, which was emphasized also in
\cite{Friess:2005tz}, is the following. One should demonstrate
that there is sufficient overlap between the off-shell deformations
$(x,y)$ of point $(a)$ and the actual on-shell
perturbations $\psi^I$. In \cite{Reall:2001ag}, it was pointed out
that a path in the family of off-shell geometries is not
related directly to an eigenfunction of the Lichnerowicz operator,
but rather some linear combination of eigenfunctions. This suggests
that when the action decreases along this path, at least one
of the eigenvalues of the Lichnerowicz operator must be negative
and therefore an actual on-shell instability should exist. It is not immediately
obvious, however, that the converse is also true.
To complete the proof one should demonstrate that the
off-shell deformations and the actual on-shell perturbations
cover the same linear space.
\end{enumerate}

Before addressing each of the above points in the case of the
smeared D$p$-branes \eqref{metD0a} it will be useful to recall how
the above points facilitate the connection between the
thermodynamics and the classical stability analysis. First of all,
since the black branes \eqref{metD0a} are actual solutions of the
Einstein equations of motion they extremize the action $I$ at
special points $(x(T,\nu),y(T,\nu))$ of the two-parameter family.
This implies the vanishing of the first derivatives
\begin{equation}
\label{aag} \bigg(\frac{\pa I}{\pa x}\bigg)_{T,\nu}=0~, ~ ~
\bigg(\frac{\pa I}{\pa y}\bigg)_{T,\nu}=0~,
\end{equation}
or equivalently the standard equations
\begin{equation}
\label{aah}
T=\bigg(\frac{\pa M}{\pa
S}\bigg)_Q\bigg|_{(x,y)=(x(T,\nu),y(T,\nu))}~, ~ ~
\nu=\bigg(\frac{\pa M}{\pa
Q}\bigg)_S\bigg|_{(x,y)=(x(T,\nu),y(T,\nu))} ~.
\end{equation}
The quadratic perturbation of the action $I$ along the surface
parameterized by $(x,y)$ involves the Hessian matrix
\begin{equation}
\label{aai}
\matrbig{\frac{\pa ^2 I}{\pa x^2}}{\frac{\pa^2
I}{\pa x \pa y}}{\frac{\pa^2
I}{\pa x \pa y}}{\frac{\pa ^2 I}{\pa y^2}}=
\matrbig{\frac{\pa T}{\pa x}}{\frac{\pa \nu}{\pa x}}{\frac{\pa
T}{\pa y}}{\frac{\pa \nu}{\pa y}}
 \matrbig{\left(\frac{\partial S}{\partial
T}\right)_\nu}{\left(\frac{
\partial Q}{\partial T}\right)_\nu}{\left(\frac{\partial S}{\partial \nu}
\right)_T}{\left(\frac{\partial Q}{\partial \nu}\right)_T} {
\matrbig{\frac{\pa T}{\pa x}}{\frac{\pa \nu}{\pa x}}{\frac{\pa
T}{\pa y}}{\frac{\pa \nu}{\pa y}} }^{\rm T} \ ,
\end{equation}
which can be re-written more compactly as a matrix equation
\begin{equation}
\label{aaj} (I_{(2)})_{ab} = -M_a{}^{\alpha} (H_{G})_{\alpha
\beta} M^{\beta}{}_b ~.
\end{equation}
In this expression $M_a^{\alpha}$ is the Jacobian of the transformation from
the variables $(x,y)$ to $(T,\nu)$ and $H_G$ is the Hessian of the Gibbs free
energy. Assuming $M$ to be an invertible matrix, equation
(\ref{aaj}) allows for a direct connection between classical
stability and local thermodynamic stability. Indeed, given
the validity of points ($b$) and $(c)$ above,
classical stability requires $I_{(2)}$ to
have positive eigenvalues and
through (\ref{aaj}) this requirement translates to $H_G$ being
negative definite. We have already argued that this is equivalent
to the conditions \eqref{grand}.

We now proceed to demonstrate points $(a)$-$(c)$ above. The first
point $(a)$ requires the construction of a two-parameter family of
Euclidean backgrounds for smeared D$p$-branes that satisfies
(\ref{aaf}). The detailed construction appears in Appendix
\ref{app:fam} and boils down to the following facts. One can show
that there is at least one two-parameter family of Euclidean
backgrounds satisfying all the Hamiltonian constraints. These
constraints include the vanishing of the Hamiltonian, the
vanishing of ten momenta and the Gauss constraint. The requirement
(\ref{aaf}) is an automatic consequence of the validity of these
constraints \cite{Hawking:1996fd,Hawking:1995ap}. The explicit
form of the family arises from the general spherically symmetric
ansatz
\begin{equation}
\label{aaka}
\begin{array}{c}
\ds ds^2=U(r)d\tau^2+V^{-1}(r)dr^2+S(r)\bigg(\sum_{i=1}^p dx_i^2
+dz^2\bigg)+ R(r) r^2 d\Omega^2_{7-p} ~,
\\[4mm] \ds
e^{2\phi}=H(r)^{\frac{3-p}{2}}~, ~ ~ F_{(p+2)}=F(r) dt \wedge dr
\wedge d x_1 \wedge \cdots \wedge dx_p ~,
\end{array}
\end{equation}
in the following way. The functions $U,V,S,R,H$ and $F$ have to be
chosen so that the Hamiltonian constraints are satisfied and so
that for specific points in the family we get the on-shell
solutions \eqref{metD0a} in the Einstein frame. In addition,
certain boundary conditions must be imposed. As usual for gravity
in the canonical ensemble, boundary conditions are imposed at an
asymptotic boundary at $r=r_B$, where the Euclidean time direction
is compactified at a radius $\beta=1/T$. The functions $U(r)$ and
$V(r)$ are both vanishing at the horizon $r=r_0$ and the relation
between $r_0$ and $T$ follows from the condition of regularity at
$r=r_0$, which in general takes the form
\begin{equation}
\label{aan} \sqrt{U'(r_0)V'(r_0)}=\frac{4\pi}{\beta}~.
\end{equation}

The two-parameter family of Appendix \ref{app:fam} is based on the
special ansatz
\begin{equation}
\label{aana} S(r)=R(r)~, ~ ~ V(r)=-H(r)^{\frac{3-p}{4}} U(r)
\spa
F(r)=a~ H(r)^{\frac{p-3}{2}} S(r)^{p-4} r^{p-7}
~,
\end{equation}
where $a$ is some integration constant that can be fixed by
imposing the appropriate boundary conditions. With this ansatz one
can satisfy the Hamiltonian constraints by suitably expressing the
functions $U,V,R$ and $F$ in terms of $H$ and $S$, which remain
free. This gives rise to a two-parameter family of black brane
backgrounds controlled by the free functions $H$ and $S$. In this
manner, the family appearing in Appendix \ref{app:fam} is a
generalization of the one-parameter family of Euclidean geometries
presented in \cite{Reall:2001ag}, following previous work in
Refs.\ \cite{Whiting:1988qr,Prestidge:1999uq}. In
\cite{Reall:2001ag} one could satisfy the Hamiltonian constraints
with a set of backgrounds depending on one free function, here we
can satisfy them with two free functions.

The choice of the functions $H$ and $S$ is arbitrary up to certain
boundary conditions. For the metric they are as follows. First, we
keep the boundary conditions at the boundary $r=r_B$ invariant.
This gives a single temperature $T$ for the whole family. Also, to
keep the topology invariant the functions $U$ and $V$ continue to
vanish at $r=r_0$ and by regularity at $r=r_0$ we demand that
their derivatives satisfy (\ref{aan}). For $r>r_0$, $U(r)$ should
remain positive. Similarly for the gauge field, in order to obtain
a family with a single chemical potential $\nu$, we want to keep
the value of the gauge field at $r=r_0$ fixed. Since we can
satisfy all of these conditions on the actual on-shell solutions
\eqref{metD0a}, we can also satisfy them for the above off-shell
family, at least infinitesimally close to the on-shell points.
This is enough for the purpose of proving the
CSC.\footnote{Similar statements apply to smeared black branes in
the near-extremal limit.}

Note that we can deduce a completely different off-shell family of
Euclidean black branes by applying the boost/U-duality procedure
of Section \ref{sec:boost} on an eleven-dimensional neutral
off-shell family of the same form as in \cite{Reall:2001ag}. There
are several problems with such a family. First of all, it would be
a one-parameter family, since we start with a one-parameter family
in eleven dimensions. Secondly, it would not satisfy all the
Hamiltonian constraints of type II supergravity, but only part of
them - more precisely, a linear combination of the Gauss
constraint and the Hamiltonian. This fact alone would be enough to
show that this family has the property (\ref{aaf}),\footnote{To
demonstrate (\ref{aaf}) we need to show that the Hamiltonian of
the system receives contributions from boundary terms only. This
is true for the off-shell family coming from the boost/U-duality
procedure, because the Hamiltonian receives only boundary
contributions in eleven dimensions. This continues to hold through
the boost/U-duality procedure. Indeed, the specific family is
time-independent and hence the Hamiltonian is proportional to the
Lagrangian, which is U-duality invariant.} but this would not be
fully satisfactory for the purposes of the CSC proof.

The boost/U-duality procedure is much more useful in demonstrating
the second point $(b)$. The actual form of the on-shell
perturbations will be discussed in the next section, but we are
already in position to argue in favor of the validity of $(b)$.
{}From the GL analysis it is clear that the norm of the on-shell
perturbations in eleven dimensions is positive. The norm is
clearly invariant under rotations and boosts and should remain
positive under U-duality, which is of course a symmetry of the
theory. Thus, the positivity of the norm is guaranteed.

The final point $(c)$ is slightly more subtle. Even in the
unsmeared case of \cite{Reall:2001ag} this is a point that has not
been shown rigorously. We do not have anything new to add on this
point for the present case of smeared D$p$-branes, but in general
it is natural to expect that there will be sufficient overlap
between the eigenfunctions of the Lichnerowicz operator and the
family of off-shell deformations for systems that are specified
uniquely by the full set of conserved charges. For example,
systems with scalar hair do not fall into this category. This
crucial point was put forward in a very recent paper by Friess,
Gubser and Mitra \cite{Friess:2005zp}, who found explicit
counter-examples to the CSC. In the context of the proof outlined
in the present paper, it is clear that in these examples the
validity of point $(c)$ breaks down.

\subsubsection*{Comments on the general proof of the CSC}

The correlated stability conjecture is expected to hold for more
general gravitational systems with $n$ chemical potentials $\nu_i$
and $n$ conjugate charges $Q_i$ ($n$ is an arbitrary positive
integer). The basic elements of Reall's proof in this general case have been
sketched recently in \cite{Friess:2005tz} and are a straightforward generalization
of the points ($a$), ($b$) and ($c$) appearing above.
More specifically, in the grand canonical ensemble one
has to show the existence of an $(n+1)$-parameter family of
off-shell deformations with Euclidean action
\begin{equation}
\label{aap} I(x_1,...,x_{n+1};\beta,\nu_1,\ldots,\nu_n)=\beta E(x_1,...,x_{n+1})-
S(x_1,...,x_{n+1})-\beta \sum_{j=1}^n \nu_j
Q_j(x_1,...,x_{n+1}) ~.
\end{equation}
and then verify the validity of the points ($b$) and ($c$) above.

As mentioned in Section \ref{sec:conj}, more general examples
involving bound states of the form D$(p-2)$-D$p$, F1-D$p$ and
D$(p-4)$-D$p$ have been discussed recently in
\cite{Gubser:2004dr,Ross:2005vh,Friess:2005tz}. The choice of
ensemble in these cases is determined by the criteria presented
above.

\subsubsection*{The CSC in the presence of compact directions}

We would like to conclude this section with a few words about the
validity of the CSC in the presence of compact directions.
Clearly, this is a situation where the CSC is expected to break
down. This can be seen explicitly, for example, in neutral uniform
black strings on a transverse circle \eqref{bsmet}. The specific
heat of the uniform branch is always negative and local
thermodynamic stability would suggest the presence of a classical
instability for black strings of any mass. Instead, the explicit
classical stability analysis by Gregory and Laflamme (see below)
exhibits a marginally stable threshold mode at a specific critical
GL mass $\mu_{GL}$ and a classical instability below the critical
mass. The uniform solution is stable above $\mu_{GL}$. It is not
difficult to see what goes wrong with the CSC proof of
\cite{Reall:2001ag} in this case. The arguments based on
thermodynamics predict the existence of a negative eigenvalue mode
$\psi^I$ for the dimensionally reduced Lichnerowicz operator, but
the actual static GL mode is $\Psi^I=Re(\psi^I e^{ikz})$. For a
transverse circle parameterized by $z$ the momentum $k$ is
quantized in integer multiples of $2\pi/L$ ($L$ being the
circumference of the $S^1$) and the static mode can exist only at
specific values of the ratio $r_0/L$ (see \eqref{bsmet} for the
definition of $r_0$), or equivalently at specific values of the
mass. The moral of this example is the following. In general, one
has to fit the GL static mode into the compact directions and this
is not automatic. Despite this complication, there is still some
practical value in the CSC in the presence of compact directions.
Although it does not fix the precise value of the critical point,
it predicts that a static mode exists whenever it can fit into the
compact directions. It is in this spirit that we use the CSC later
on in the context of non-extremal and near-extremal branes smeared
on a transverse circle.

\section{The Gregory-Laflamme mode for smeared branes\label{sec:mode}}

In this section we show that the boost/U-duality transformation reviewed in
Section \ref{sec:boost} can be used to transform the time-dependent
unstable mode of a neutral black string, found by Gregory and Laflamme,
to a time-dependent unstable mode of non- and
near-extremal smeared D$p$-branes. As explained below, this section relies
crucially on an argument of \cite{Aharony:2004ig}.

\subsection{Neutral black strings}

As discovered by Gregory and Laflamme
\cite{Gregory:1993vy,Gregory:1994bj}, the black string in
$(10-p)$-dimensional Minkowski-space \eqref{bsmet} is classically
unstable under linear perturbations $g_{\mu \nu} + h_{\mu\nu}$ of
the form
\begin{equation}
\label{GLmode} \begin{array}{c} \ds h_{\mu \nu} = \mbox{Re}
\left\{ \exp \left( \frac{\Omega t}{r_0} + i \frac{kz}{r_0}
\right) P_{\mu \nu} \right\} \ ,
\\[4mm] \ds
P_{tt} = - f \psi \spa P_{tr} = \eta \spa P_{rr} = f^{-1} \chi
\spa P_{\rm sphere} = r^2\kappa \ ,
\end{array}
\end{equation}
where $\psi$, $\eta$, $\chi$ and $\kappa$ are functions for which the
perturbation $h_{\mu\nu}$ solves the linearized Einstein equations.
Note that we define $\psi$, $\eta$, $\chi$ and $\kappa$ to be functions of the
parameter $x = rk/r_0$, which is well defined under the
near-extremal and extremal limits in the charged case below.
The gauge conditions on $h_{\mu\nu}$ are
the tracelessness condition \eqref{traceeq} and the transversality
conditions \eqref{transeqs}. The Einstein equations reduce
to the four independent equations
\eqref{E1}-\eqref{E4} that appear in Appendix \ref{app:mode}.

\begin{figure}
\label{figGL}
\begin{center}
\resizebox{10cm}{6cm}{\includegraphics{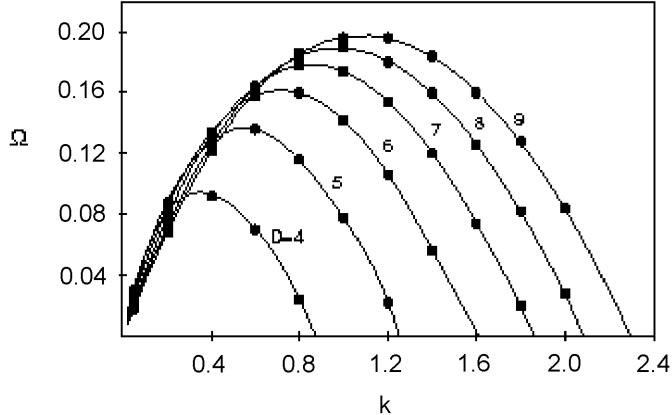}}
\end{center}
\caption{The Gregory-Laflamme $\Omega(k)$ curve in $D+1=10-p$
dimensions. Reprinted from Ref.~\cite{Gregory:1993vy}.}
\label{fig:branes}
\end{figure}

Combining the gauge conditions with the linearized Einstein
equations of motion one can derive a single second order
differential equation on $\psi$
\begin{equation}
\label{psieq} \psi''(x) +{\cal Q}_p(x) \psi'(x) +{\cal P}_p(x) \psi(x) = 0
~.
\end{equation}
The explicit form of the $p$-dependent rational functions
${\cal Q}_p(x)$ and ${\cal P}_p(x)$ appears in Appendix \ref{app:mode}.
This equation depends only on $k$, $\Omega$ and the variable $x$.
Then one gets a curve of the possible values of $\Omega$ and $k$
for which this equation has a solution. We plot a sketch of
this curve found numerically in
\cite{Gregory:1993vy,Gregory:1994bj} in Figure \ref{figGL}.
Notice that the well-known threshold GL mode appears at
the critical value $k=k_c$ for which $\Omega=0$.

\subsection{Non-extremal smeared D$p$-branes}

In Section \ref{sec:boost} we reviewed the boost/U-duality
transformation of \cite{Harmark:2004ws} that takes a neutral black
string to a D$p$-brane solution smeared on a transverse direction.
In this section we describe the transformation that takes the
unstable GL mode of the neutral black string to an unstable mode
of the non-extremal D$p$-brane solution smeared on a transverse
circle.

The transformation should convert the neutral black string
solution \eqref{bsmet11} to the smeared D$p$-brane solution
\eqref{metD0a}. If we try naively to apply the boost/U-duality
transformation reviewed in Section \ref{sec:boost} to the GL mode
\eqref{GLmode} we find a non-normalizable exponential dependence
on $y$ (the critical case $k=k_c$ is an exception). As was pointed
out in \cite{Aharony:2004ig}, to avoid this problem we may
consider a complex metric perturbation, and try to find a complex
transformation that i) Gets rid of the exponential $y$ dependence,
ii) Still has the same effect on the zeroth order part of the
metric, $i.e.$ on the neutral black string metric. At the end one
can take the real part of the transformed perturbation. Note that
this last step works only because of the linearity of the
perturbed Einstein equations of motion.

The precise form of the transformation we look is as follows.
First define the coordinates $\tilde{y}$ and $z'$ by the following
complex rotation of the $y$ and $z$ coordinates
\begin{equation}
\label{complrot} \vecto{y'}{\tilde{z}} = M \vecto{y}{z} \spa M =
\matrto{\cosh w}{-i \sinh w }{i \sinh w}{\cosh w} \ .
\end{equation}
Notice that the neutral black string metric \eqref{bsmet11} embedded in eleven
dimensional gravity is invariant under \eqref{complrot} since
$(dy')^2 + (d\tilde{z})^2 = dy^2 + dz^2$. The transformation
\eqref{complrot} is then supplemented by a boost that takes the $(t,y')$
coordinates to the $(\tilde{t},\tilde{y})$ coordinates given by
\begin{equation}
\vecto{\tilde{t}}{\tilde{y}} = \Lambda \vecto{t}{y'} \spa \Lambda
= \matrto{\cosh \alpha }{\sinh \alpha}{\sinh \alpha}{\cosh \alpha}
~.
\end{equation}
These rotations are such that
\begin{equation}
\label{ktilde} \sinh w = \frac{\Omega}{k} \tanh \alpha \spa
\tilde{k} = k\cosh w \spa \tilde{\Omega} = \frac{\Omega}{\cosh
\alpha}
~.
\end{equation}
Then we can write the real part of the boosted perturbation of the
eleven dimensional metric \eqref{boomet} as
\begin{equation}
\label{boostmodeI} \tilde{h}_{\mu \nu} = \mbox{Re} \left\{ \exp
\left( \frac{\tilde{\Omega} t}{r_0} + i \frac{\tilde{k}z}{r_0}
\right) \tilde{P}_{\mu \nu} \right\}
~.
\end{equation}
Using the relation $\tilde{P} = \Lambda^{-1} M^{-1} P (M^{-1})^T
(\Lambda^{-1})^T$ we find
\begin{equation}
\label{boostmodeII}
\begin{array}{c} \ds \tilde{P}_{tt} = - f \psi
\cosh^2 \alpha \spa \tilde{P}_{yy} = - f \psi \sinh^2 \alpha \spa
\tilde{P}_{ty} = f \psi \sinh \alpha \cosh \alpha \ ,
\\[3mm] \ds
\tilde{P}_{tr} = \eta \cosh \alpha \spa \tilde{P}_{yr} = - \eta
\sinh \alpha \spa \tilde{P}_{rr} = f^{-1} \chi \spa \tilde{P}_{\rm
sphere} = r^2 \kappa ~,
\end{array}
\end{equation}
where we have made the relabellings $\tilde{t} \rightarrow t$ and
$\tilde{y} \rightarrow y$. Notice that the complex rotation
\eqref{complrot} precisely ensures that the exponential factor
does not have a $y$-dependence. We can therefore apply the same
U-duality transformations on
\eqref{boostmodeI}-\eqref{boostmodeII} as on the boosted neutral
black string. After these U-duality transformations, consisting of
one S-duality and $p$ T-dualities, we conclude that the perturbed
non-extremal D$p$-brane smeared on a transverse direction
can be written as%
\footnote{As usual, in the special case of D3-branes the gauge field
strength is self dual, so that we have
$F_5=(dA_4+\star dA_4)/\sqrt{2}$.}
\begin{equation}
\label{nonmode}
\begin{array}{c} \ds
\begin{array}{rcl}
ds^2 &=& \ds H_{\rm c}^{-1/2} \left[ - f_{\rm c} dt^2 +
\sum_{i=1}^p dx_i^2 + 2 \eta \cosh \alpha {\cal E} dt dr \right]
\\[5mm] && \ds + H_{\rm
c}^{1/2} \Big[ f^{-1} (1+\chi {\cal E}) dr^2 + dz^2 +    r^2
(1+\kappa {\cal E}) d\Omega_{7-p}^2 \Big] \ ,
\end{array}
\\[12mm] \ds e^{2\phi} = H_{\rm c}^{(3-p)/2} \spa A_{01\cdots p} = \coth
\alpha (H_{\rm c}^{-1} - 1) \spa A_{r1\cdots p} = - H^{-1} \eta
\sinh \alpha {\cal E} \ ,
\\[3mm] \ds f = 1 - \frac{r_0^{6-p}}{r^{6-p}} \spa
H=1 + \sinh^2 \alpha ( 1 - f ) \spa f_{\rm c} \equiv f (1 + \psi
{\cal E} ) \spa H_{\rm c} \equiv 1 + \sinh^2 \alpha ( 1 - f_{\rm
c} ) \ ,
\\[3mm] \ds
{\cal E} = \cos \left(\tilde{k} r_0^{-1} z \right) \exp \left(
\tilde{\Omega} r_0^{-1} t \right) \ .
\end{array}
\end{equation}
The perturbed D$p$-brane background appears here
in one compact expression. From this one can easily deduce the
explicit form of the perturbation by expanding to first
order. The unstable mode appearing in \cite{Aharony:2004ig}
is equivalent to \eqref{nonmode} for $p=0$, though written in a different gauge.

Note that the functions $\psi$, $\eta$, $\chi$ and $\kappa$ are
still solutions to Eqs.~\eqref{E1}-\eqref{E4} with $f$ given by
\eqref{fkx}. In particular $\psi$ is a solution of
Eq.~\eqref{psieq}. All of these functions depend on the
variable $x = rk/r_0$ just as in the case of the neutral black string
perturbation.

Furthermore, we see from \eqref{ktilde} that $\tilde{k}^2 = k^2 +
\Omega^2 \tanh^2 \alpha$ and $\tilde{\Omega} = \Omega /\cosh
\alpha$. Hence, we can obtain $\tilde{\Omega}$ as function of
$\tilde{k}$ by using the functional dependence $\Omega(k)$ for the
neutral black string (as sketched for $p=0,...,5$ in Figure
\ref{figGL}). We note that the critical point $(k,\Omega)=(k_c,0)$
is mapped to the critical point $(\tilde{k},\tilde{\Omega})=(k_c ,
0)$, which corresponds to the marginal mode of the D$p$-brane with
a transverse direction. This marginal mode appears at the origin
of the non-uniform phase of non-extremal D$p$-branes with a
transverse direction \cite{Harmark:2004ws}. From Figure
\ref{figGL} we also deduce that $\tilde k<k_c$ for any $k<k_c$ and
the function $\tilde \Omega(\tilde k)$ never exhibits
time-dependent modes with wavelength smaller than the critical
one.

The existence of the perturbation \eqref{nonmode}
proves explicitly that a non-extremal D$p$-brane with a transverse
flat direction is always unstable as a gravitational background.
This result meshes nicely with the unstable thermodynamics
in \eqref{cq2}, \eqref{cq3} and confirms the validity of the
CSC if we consider it in the grand canonical ensemble.
In contrast to this, a D$(p+1)$-brane wrapped on the T-dual $z$-circle
is known to be classically stable in supergravity close to extremality.
This fact is in agreement with the CSC if we consider it
in the canonical ensemble (see \eqref{cq2}).

\subsection{Near-extremal limit}
\label{sec:neGLmode}

We now consider the near-extremal limit \eqref{nelim} of the
unstable mode \eqref{nonmode} for non-extremal D$p$-branes with a
transverse direction. The near-extremal limit of the unperturbed
non-extremal solutions \eqref{metD0a} has been discussed already
in Section \ref{sec:prem}, giving the near-extremal D$p$-brane
background \eqref{NEbrane}.

Applying the limit \eqref{nelim} on the background \eqref{nonmode}
gives the following perturbation of the near-extremal D$p$-brane
\eqref{NEbrane}
\begin{equation}
\label{nearmode}
\begin{array}{c}
\begin{array}{rcl}
l_s^{-2} ds^2 &=& \ds \hat{H}_{\rm c}^{-1/2} \left[ - f_{\rm c}
dt^2 + \sum_{i=1}^p dx_i^2 + 2 \eta \sqrt{K} u_0^{p/2-3} {\cal E} dt du
\right]
\\[5mm]
&& \ds + \hat{H}_{\rm c}^{1/2} \Big[ f^{-1} (1+\chi {\cal E}) du^2
+ d\hat{z}^2 + u^2 (1+\kappa {\cal E}) d\Omega_{7-p}^2 \Big] \ ,
\end{array}
\\[12mm] \ds
e^{2\phi} = \hat{H}_{\rm c}^{(3-p)/2} \spa A_{01\cdots p} =
\hat{H}_{\rm c}^{-1} \spa A_{r1\cdots p} = - u^{6-p} K^{-1/2}
u_0^{p/2-3} \eta {\cal E} \ ,
\\[3mm] \ds f = 1 - \frac{u_0^{6-p}}{u^{6-p}} \spa
f_{\rm c} = f ( 1 + \psi {\cal E} ) \spa \hat{H}_{\rm c} =
\frac{K}{u^{6-p}} \left[ 1 - \left( \frac{u^{6-p}}{u_0^{6-p}} - 1
\right) \psi {\cal E} \right] \ ,
\\[4mm] \ds
{\cal E} = \cos \left( \sqrt{k^2+\Omega^2} \frac{ \hat{z}}{u_0}
\right) \exp \left( u_0^{2-\frac{1}{2}p} K^{-1/2} \Omega t \right)
\ .
\end{array}
\end{equation}
We note that the near-extremal limit \eqref{nelim} keeps $k$ and
$\Omega$ fixed, and moreover keeps $\psi$, $\eta$, $\chi$ and
$\kappa$ fixed as functions of the variable $x$ which is now given
by
\begin{equation}
x = \frac{u k}{u_0} \ .
\end{equation}
Therefore, the functions $\psi$, $\eta$, $\chi$ and $\kappa$ are
still solutions of the Eqs.~\eqref{E1}-\eqref{E4} with $f$ given by
\eqref{fkx}. In particular $\psi$ is still a solution of
Eq.~\eqref{psieq}.

We have thus shown that the near-extremal limit of the
time-dependent mode of smeared branes is well-defined and hence
that near-extremal smeared D$p$-branes are classically unstable.
Then according to the CSC this means that near-extremal
D$p$-branes should be thermodynamically unstable in the grand
canonical ensemble. However, as noted in Section \ref{sec:thermo},
we do not have a priori a definition of this ensemble for
near-extremal branes, since it seems that the charge cannot vary
anymore. We will resolve this puzzle in Section \ref{sec:puz},
where we will show that there does exist an appropriate definition
of this ensemble in the near-extremal limit. We will see that in
this ensemble near-extremal smeared D$p$-branes are indeed
thermodynamically unstable, in accordance with the above result
and the CSC.

\section{Connection to marginal modes of extremal smeared branes}
\label{sec:ext}

In this section we find a connection between the GL mode of
charged smeared branes described in Section \ref{sec:mode} and
the marginal modes of extremal smeared branes.

\subsection{Marginal modes for extremal smeared branes}
\label{sec:margmodes}

In general, the solutions for extremal D$p$-branes distributed
along a single flat direction $z$ is given as
\begin{equation}
\label{extrback}
\begin{array}{c} \ds
ds^2 = H^{-1/2} \left( - dt^2 + \sum_{i=1}^p dx_i^2 \right) +
H^{1/2} \left( dr^2 + dz^2 + r^2 d\Omega_{7-p}^2 \right) \ ,
\\[5mm] \ds
e^{2\phi} = H^{(3-p)/2} \spa A_{01\cdots p} = H^{-1} - 1 \ ,
\end{array}
\end{equation}
where the harmonic function $H(r,z)$ obeys
\begin{equation}
\left( \partial_r^2 + \frac{7-p}{r} \partial_r + \partial_z^2
\right) H (r,z) =0 \ ,
\end{equation}
away from the source distribution. Given the appropriate boundary conditions
the general solution is
\begin{equation}
H(r,z) = 1 + \int_{-\infty}^{\infty} dz'
\frac{\rho(z')}{(r^2+(z-z')^2)^{(7-p)/2}} \ ,
\end{equation}
where the charge distribution function $\rho(z)$ is arbitrary.
Now, the extremal smeared D$p$-brane case corresponds obviously to
$\rho$ being constant. But, it is clearly possible to add to the
constant charge distribution a single mode of any wave-number $q$,
in which case the harmonic function takes the form
\begin{equation}
\label{Hm} H = 1 + \frac{Kl_s^{8-2p}}{r^{6-p}} + m(qr) \cos( q z)
\ ,
\end{equation}
where $K$ is defined in \eqref{Kdef} and the function $m(q r)$
solves
\begin{equation}
\label{meq} m''(y) + \frac{7-p}{y} m'(y) - m(y) = 0 \ .
\end{equation}
The solutions of Eq.~\eqref{meq} are of the form $m(y) \propto
y^{-(6-p)/2)} {\cal{K}}_{(6-p)/2} (y)$, ${\cal{K}}_s(y)$ being the
modified Bessel function of the second kind. The modes given by
\eqref{Hm} are marginal in the sense that the modified solution
solves the supergravity equations. In this sense the extremal
smeared branes have marginal modes of any wave-length.

It is interesting to consider what happens to such marginal modes
if one perturbs the extremal brane so that it gets a
temperature. This is one of the things which we address in Section
\ref{sec:extrlim}. Clearly from the study of non-extremal
D$p$-branes distributed on transverse directions, one does not
expect the existence of static solutions for any charge
distribution
$\rho(z)$.%
\footnote{In fact one only knows the case of the uniform
distribution where $\rho$ is constant and the case of the
localized distribution where $\rho = c \sum_{n\in \Z} \delta (z -
nl)$ with $c$ a constant and $l$ a length.
These corresponds to the uniform and localized phase of
non-extremal branes on a circle \cite{Harmark:2004ws}.} This is
one example where gravity behaves vastly different from, say,
electromagnetism where solutions with arbitrary charge
distributions exist, although most of them are not stable when
interactions are taken into account.

Since we shall explore in the following a connection with
near-extremal branes, we need to consider briefly the near-horizon
limit \eqref{nelim} of the extremal D$p$-brane background given by
\eqref{extrback} and \eqref{Hm}. This gives
\begin{equation}
\label{extrnh}
\begin{array}{c} \ds
l_s^{-2} ds^2 = \hat{H}^{-1/2} \left( - dt^2 + \sum_{i=1}^p dx_i^2
\right) + \hat{H}^{1/2} \left( du^2 + d\hat{z}^2 + u^2
d\Omega_{7-p}^2 \right) \ ,
\\[2mm] \ds
e^{2\phi} = \hat{H}^{(3-p)/2} \spa A_{01\cdots p} = \hat{H}^{-1}
\spa \hat{H} = \frac{K}{u^{6-p}} + m(qu) \cos( q \hat{z}) \ .
\end{array}
\end{equation}
Note that the function $m(qu)$ has been rescaled appropriately,
and that $m(y)$ still obeys Eq.~\eqref{meq}.

\subsection{Extremal limit of the Gregory-Laflamme perturbation}
\label{sec:extrlim}

In this section we consider the extremal limit of the GL mode for
near-extremal smeared D$p$-branes, as found in Section
\ref{sec:neGLmode}. As we see in a moment, this gives a connection
between the marginal modes of extremal smeared branes considered
in Section \ref{sec:margmodes} and the GL modes of near-extremal
smeared branes.

Let us take the extremal limit of the perturbation of
near-extremal D$p$-branes smeared on a transverse direction, as
given by \eqref{nearmode}. The extremal limit corresponds to $u_0
\rightarrow 0$ with $u$ and $K$ kept fixed. We first remark that
since we want the argument of the cosine factor to remain finite,
we need $\sqrt{k^2+\Omega^2} / u_0$ to stay finite in the limit.
Hence, we need $k/u_0$ and $\Omega/u_0$ to remain fixed in the
limit $u_0 \rightarrow 0$. In terms of the dispersion diagram of
Figure \ref{figGL} this means that we move closer and closer to
the point $(k,\Omega)=(0,0)$ on the left part of the curve. We see
that the variable $x = uk/u_0$ defined above stays finite in the
limit. Note also that this limit necessarily has the consequence
that the time-dependent exponential factor in ${\cal E}$ in
\eqref{nearmode} disappears. One can now see from $\hat{H}_c$ in
\eqref{nearmode} that we need $\psi(x) / u_0^{6-p}$ to stay
finite. By using this in Eqs.~\eqref{traceeq}-\eqref{E4} and
demanding consistency of this system of equations, we get similar
conditions on $\eta$, $\chi$ and $\kappa$, so that in total the
extremal limit is given as
\begin{equation}
\label{extrlim} u_0 \rightarrow 0 \ \mbox{ with }\  u \ , \ K  \ ,
\ \frac{k}{u_0} \ , \ \frac{\Omega}{u_0} \ , \
\frac{\psi(x)}{u_0^{6-p}} \ , \ \frac{\eta(x)}{u_0^{6-p}} \ , \
\frac{\chi(x)}{u_0^{6-p}} \ , \ \frac{\kappa(x)}{u_0^{6-p}} \
\mbox{ kept fixed} \ .
\end{equation}
Applying the extremal limit \eqref{extrlim} to the perturbed near-extremal
solution given in \eqref{nearmode}, we get the
following background
\begin{equation}
\label{extrmode}
\begin{array}{c}
\begin{array}{rcl}
l_s^{-2} ds^2 &=& \ds \hat{H}_{\rm c}^{-1/2} \left[ dt^2 +
\sum_{i=1}^p dx_i^2  \right] + \hat{H}_{\rm c}^{1/2} \Big[  du^2 +
d\hat{z}^2 + u^2 d\Omega_{7-p}^2 \Big] \ ,
\end{array}
\\[4mm] \ds
e^{2\phi} = \hat{H}_{\rm c}^{(3-p)/2} \spa A_{01\cdots p} =
\hat{H}_{\rm c}^{-1} \spa \hat{H}_{\rm c} = \frac{K}{u^{6-p}} -
\frac{\psi}{u_0^{6-p}} {\cal E} \ ,
\\[4mm] \ds
{\cal E} = \cos \left(  \frac{\sqrt{k^2+\Omega^2}}{u_0}
\hat{z}\right) \ .
\end{array}
\end{equation}
We see that out of the four functions $\psi$, $\eta$, $\chi$ and
$\kappa$ making up the near-extremal GL mode only $\psi$ survives.
Moreover, the equation \eqref{psieq} for $\psi(x)$ becomes
\begin{equation}
\label{extrpsieq}
\psi''(x)+\frac{7-p}{x} \psi'(x)-
\bigg(1+\frac{\Omega^2}{k^2}\bigg)\psi(x)=0
~.
\end{equation}
We deduce that the background \eqref{extrmode} precisely
corresponds to the background \eqref{extrnh} describing a extremal
smeared D$p$-brane perturbed by a marginal mode with the
wave-number
\begin{equation}
q = \frac{\sqrt{k^2+\Omega^2}}{u_0} \ .
\end{equation}
In particular, one can check explicitly that the $\psi$ equation
\eqref{extrpsieq} turns into Eq.~\eqref{meq} by identifying $m
(qu) = \psi (k u / u_0)$. This means that the extremal limit of the GL
mode becomes stable as also observed for unsmeared black branes in
Ref.~\cite{Gregory:1994tw}.

It is easy to see that we can get marginal modes of any
wave-number. For small $k$ we have $\Omega = \gamma k$, where
$\gamma$ is an appropriate number. Therefore, $q =
\sqrt{1+\gamma^2} k /u_0$, and by choosing $k/u_0$ we can get any
value of $q$ that we want. We have thus shown that in the extremal
limit the GL modes of the near-extremal smeared branes precisely
become the marginal modes of extremal smeared branes. This answers
the question posed in Section \ref{sec:margmodes} of what happens
to a marginal mode of an extremal smeared brane when perturbing
the background so that it turns non-BPS: The mode will turn into
the near-extremal GL mode \eqref{nearmode}. Obviously, this has
the consequence that the brane background is classically unstable
and it will therefore decay.

Moreover, one can imagine having a charge distribution of the form
\begin{equation}
\label{rhorho} \rho = \rho_0 + \sum_{n \neq 0} \rho_n \cos( q_n
\hat{z} ) \ ,
\end{equation}
where $\rho_n$ are small compared to $\rho_0$. When we make a
non-BPS perturbation this solution involves a sum over GL modes which
is of the form
\begin{equation}
h_{\mu\nu} = \sum_{n \neq 0} h^{(n)}_{\mu\nu} (u) \cos( q_n
\hat{z} ) \exp \left( \frac{\gamma
u_0^{3-\frac{p}{2}}}{\sqrt{K}\sqrt{1+\gamma^2}} q_n t \right) \ ,
\end{equation}
for $u_0$ being small. In this way we can find a non-BPS
continuation of any extremal perturbation of the extremal smeared
branes, since we can fourier decompose the perturbation and put it
in the form \eqref{rhorho}.

It is important to note that the connection between the marginal
modes of extremal smeared branes and the GL modes of near-extremal
branes easily generalizes to a connection with GL modes of
non-extremal branes. We choose for convenience here only to
consider the connection to near-extremal branes.

The physical consequence of the connection between marginal modes
of the extremal smeared branes and the GL mode of the non- and
near-extremal smeared branes is that extremal smeared branes are
in a sense arbitrarily close to being unstable. Therefore, if one
imagined a practical application of extremal smeared branes one
would run into the problem that any disturbance of the background
which is non-BPS would cause the whole background to destabilize
and decay. Also, this feature of extremal smeared branes makes it
very unlikely that they form in any cosmologic scenario, even
though they are stable by themselves.

\section{The CSC for near-extremal branes \label{sec:puz}}

In this section, we return to the puzzle alluded to at the end of
Section \ref{sec:thermo} and further commented on at the end of
Section \ref{sec:neGLmode}. We have seen in Section
\ref{sec:neGLmode}, by applying boost/U-duality and the
near-extremal limit,  that near-extremal smeared branes are
classically unstable. According to the CSC this classical
instability should be related to thermodynamic instability in the
grand canonical ensemble, when taking the near-extremal limit. On
the other hand, we observed in Section \ref{sec:thermo} that in
the near-extremal limit we have $C_Q > 0$ and $c =0$ for
near-extremal smeared branes, suggesting marginal thermodynamic
stability. This seems to be in contradiction with the CSC.

The resolution of the puzzle is that in the near-extremal limit one
should not compute the quantity $c$ using the non-extremal
chemical potential $\nu$, but using a new chemical potential $\hat \nu$,
which can be understood roughly as the ``chemical potential above extremality".
This is in analogy to the fact that we do not compute $C_Q$ using the mass
$M$ (which goes to infinity in the near-extremal limit), but using
the energy $E$ above extremality, which is finite.

To derive the expression for $\hat \nu$, we note that the first law of
thermodynamics for non-extremal branes takes the form
\begin{equation}
\label{law1} d E = T d S + (\nu -1) d Q \ ,
\end{equation}
when written in terms of $E = M -Q$. Moreover, one can easily see
from \eqref{thermon1}, that the chemical potential $\nu$
approaches 1 in the near-extremal limit $\alpha \rightarrow
\infty$, and is not a free parameter anymore. It is therefore
natural to define the rescaled chemical potential $\hat \nu$ and
the corresponding rescaled charge ${\hat Q}$ as
\begin{equation}
\label{nuhatdef} \hat \nu \equiv \frac{1}{\gym^2} \lim_{l_s \rightarrow 0}
\frac{1}{ l_s^4} (\nu -1 )
\spa \hat Q \equiv \gym^2\lim_{l_s \rightarrow 0}  l_s^4 Q \ ,
\end{equation}
where the factor of $l_s$ has been chosen such that $\hat \nu$ is
finite in the near-extremal limit. The finite factor $\gym^2$ has
been inserted for later convenience. With the definitions
\eqref{nuhatdef}, we then find that \eqref{law1} can be written as
\begin{equation}
\label{law1p} d E = T d S + \hat \nu d \hat Q \ ,
\end{equation}
which is a well-defined differential relation in the near-extremal
limit. We have thus obtained  a new version of the first law of
thermodynamics for near-extremal branes involving an extra term
containing the rescaled chemical potential $\hat \nu$ and rescaled
charge $\hat Q$.

In further detail, using the non-extremal quantities
\eqref{thermon1} and the near-extremal limit \eqref{nelim} in the
definitions  \eqref{nuhatdef} we find for near-extremal smeared
D$p$-branes the expressions
\begin{equation}
\label{thermon5} \hat Q = (6-p)  \frac{K}{{\cal{G}}} \spa \hat \nu
= - \frac{u_0^{6-p}}{2 K} \ ,
\end{equation}
where $K$ and ${\cal{G}}$ are defined in \eqref{Kdef}, \eqref{Gdef}.

The near-extremal Gibbs free energy is then naturally defined by
$G(T,\hat \nu) = E - TS - \hat \nu \hat Q$, and
the condition for thermodynamic stability of near-extremal
smeared branes in the grand canonical ensemble follows immediately
by analogy with \eqref{grand}, namely
\begin{equation}
C_{\hat Q}=\left(\frac{\partial E}{\partial T}\right)_{\hat Q} >0
\spa \hat c=\left( \frac{\partial \hat \nu}{\partial \hat
Q}\right)_T > 0 \ .
\end{equation}
Computing these quantities for near-extremal smeared D$p$-branes,
we find that the quantity $C_{\hat Q}$ is identical to the one computed in
\eqref{cqneex}. On the other hand, for $\hat c$ we find
using \eqref{thermon5} and \eqref{thermon3} after some algebra that%
\footnote{One can compute $\hat c$ directly from the near-extremal
thermodynamics using $\hat c = \frac{ \partial ( \hat \nu,T)
}{\partial (u_0,K)} [\frac{ \partial(\hat Q,T)}{\partial
(u_0,K)}]^{-1}$. Alternatively, one can show from \eqref{nuhatdef}
that $\hat c = \lim l_s^{-8} g_{\rm YM}^{-4} c $. We obtain the same result
by using in this expression
the non-extremal relation \eqref{cqnex} for $c$ and then taking the
near-extremal limit.}
\begin{equation}
\hat c=- \frac{{\cal{G}}}{g_{\rm YM}^4 K^2}
\frac{u_0^{6-p}}{(4-p)(6-p)} \ ,
\end{equation}
which is finite and negative for $p \leq 3$.
As a consequence we find that near-extremal smeared D$p$-branes with $p\leq 3$
are thermodynamically unstable in the near-extremal
grand canonical ensemble defined above,
in agreement with  the classical stability analysis and the CSC.

\subsubsection*{Interpretation in the dual gauge theory}

Using \eqref{Kdef}, \eqref{Gdef} in \eqref{thermon5} we can
express $\hat \nu$ and $\hat Q$ in gauge theory variables. The
result is
\begin{equation}
\hat Q =\frac{ \bar L V_p}{(2\pi)^2} N \spa \hat \nu = -
\frac{6-p}{2} \Omega_{7-p} (2\pi)^{2p-7} \frac{u_0^{6-p}}{\gym^4
N}\ .
\end{equation}
These quantities have a natural interpretation in the dual gauge
theory. In the presence of the spatial circle the gauge theory
possesses a non-trivial spatial Wilson loop. $\hat Q$ is naturally
related to the number of eigenvalues of this operator and $\hat
\nu$ is the corresponding chemical potential. Hence, the GL
instabilities that we found previously within supergravity
translate in gauge theory into a new set of phase transitions
parametrized by the Wilson loop around the spatial circle. A
recent discussion of these transitions in 0+1 and 1+1 dimensions
appeared in \cite{Aharony:2004ig} both at weak and strong
coupling.

In the present work we have seen that a near-extremal D$p$-brane
smeared on a transverse circle exhibits a momentum mode
instability when the radius of the circle becomes sufficiently
large. The smeared D$p$-brane is T-dual to a D$(p+1)$-brane on
which the dual gauge theory lives. In the T-dual picture the
near-extremal D$(p+1)$-brane exhibits a winding mode instability
along a longitudinal circle when the T-dual radius is sufficiently
small. {}From the string theory point of view the condensation of
the unstable winding mode will be a dynamical process from the
usual D$(p+1)$-brane phase to a non-trivial D$(p+1)$-brane phase
which is T-dual to the localized phase of the D$p$-brane.

We can examine what this means on the gauge theory side. For
example when considering smeared D2-branes, the above picture
 suggests
that by perturbing $N=4$ $D=4$ SYM on a spatial circle for small
temperatures with a particular type of relevant operator the
theory destabilizes and eventually settles down to a phase in
which it behaves approximately as the 2+1 dimensional SYM theory
obtained by dimensional reduction of the $N=4$ $D=4$ SYM theory.
Similar statements apply to gauge theories on the world-volume of
other D$p$-branes. It would be interesting to examine this type of
process more closely.

\section{Conclusions \label{sec:con}}

In this paper we considered non-extremal and near-extremal
D$p$-branes smeared on a transverse direction. These branes exhibit
GL instabilities, which can be obtained most easily from GL
instabilities of neutral branes by a suitable boost/U-duality map.
We showed explicitly that the unstable modes persist in the
near-extremal limit and reduce to the more standard marginal modes
of the extremal smeared branes, when we further take the extremal
limit. This picture meshes nicely with the picture suggested by a
local thermodynamic stability analysis when we apply the
correlated stability conjecture in the grandcanonical ensemble. As
we saw, a natural definition of such an ensemble exists also in
the near-extremal limit. Moreover, for this class of smeared
D$p$-branes we provided a proof of the CSC by extending previous
arguments of Reall \cite{Reall:2001ag}.

There is a number of further interesting issues related to the
above circle of ideas. One of them has to do per se with the CSC.
So far, there has been no general and complete proof of this
conjecture. It would be interesting to generalize the arguments
presented here to a wider class of systems, for example to systems
of bound states. In general, the CSC appears to be such a natural
conjecture that one may wonder whether it is possible to find a
proof based on a more universal argument (see \cite{Buchel:2005nt}
for work in this direction). In cases where the CSC is violated
\cite{Friess:2005zp} it would be interesting to examine possible
ways in which it can be restricted or revised accordingly.

In this note we treated standard unstable GL modes with
non-vanishing momentum and zero winding. A T-duality
transformation converts the smeared brane into a wrapped brane. In
string theory, the wrapping brane continues to be unstable, but
now the unstable mode has non-vanishing winding. The winding state
in question is expected to become tachyonic when the (T-dual)
radius is smaller than the critical value. It would be very
interesting to explore situations in string theory where the
presence and dynamics of this mode can be analyzed explicitly.%
\footnote{For related recent work on this see
\cite{Adams:2005rb,Horowitz:2005vp}.}

Because of the AdS/CFT correspondence each of the above phenomena
has some counterpart in the dual gauge theory. In particular, the
GL instabilities translate into a new set of phase transitions
with order parameters associated to the Wilson loops around the
non-trivial cycles of the spacetime of the gauge theory. For a
recent examination of this type of transitions in 0+1 and 1+1
dimensions see \cite{Aharony:2004ig,Aharony:2005ew}. The rescaled
``charge'' $\hat Q$ is naturally related to the number of
eigenvalues of the spatial Wilson loop observables mentioned just
above and the near-extremal quantity $\hat \nu$ is just the
corresponding chemical potential. In gravity the phase transition
to a localized phase is a dynamical process from a black string
phase to a black hole phase. It would be interesting to obtain a
better understanding of the dynamics of this process in gauge
theory.


\section*{Acknowledgments}

We thank O. Aharony and E. Lozano-Tellechea  for useful
discussions and R. Gregory for permission to reprint Figure
\ref{figGL}. Work partially supported by the European Community's
Human Potential Programme under contract MRTN-CT-2004-005104
`Constituents, fundamental forces and symmetries of the universe'.

\begin{appendix}

\section{The Gregory-Laflamme mode}
\label{app:mode}

We consider here the GL instability
\cite{Gregory:1993vy,Gregory:1994bj} for the metric \eqref{bsmet}
corresponding to a uniform black string in $10-p$
dimensions.%
\footnote{See \cite{Gross:1982cv,Prestidge:1999uq,Kol:2004pn} for
related work on this.}
The instability is given by the metric
perturbation $h_{\mu\nu}$ in the sense that $g_{\mu\nu} +
h_{\mu\nu}$ is the black string metric plus perturbation. The
perturbation $h_{\mu\nu}$ appears in \eqref{GLmode}. In
\eqref{GLmode} $\psi$, $\eta$, $\chi$ and $\kappa$ are all
functions of the variable
\begin{equation}
x = \frac{rk}{r_0} \ ,
\end{equation}
and this is the variable that we take derivatives with respect to.
The function $f$ is given in terms of $x$ by
\begin{equation}
\label{fkx} f = 1 - \frac{k^{6-p}}{x^{6-p}} \ .
\end{equation}
The tracelessness conditions are
\begin{equation}
\label{traceeq}
 (7-p) \kappa + \chi + \psi = 0 \ .
\end{equation}
The transversality condition is
\begin{equation}
\label{transeqs}
\begin{array}{c} \ds
\frac{\Omega}{k} x \psi + (6-p)(1-f) \eta + f \left((7-p) \eta + x
\eta' \right)=0 \ ,
\\[3mm] \ds
-2 \frac{\Omega}{k} x  \eta+(6-p)(1-f)(\chi-\psi)+2 f \left((8-p)
\chi+\psi+ x \chi'\right)=0 \ .
\end{array}
\end{equation}
The four independent Einstein equations take the form
\begin{eqnarray}
& \ds \label{E1} \begin{array}{l} \ds x \frac{\Omega}{k} \psi +
(6-p)(1-f) \eta +\frac{k}{\Omega} f^2 \big(
(7-p)\psi'+x\psi''\big)
\\[2mm] \ds
+f \left[ 2(7-p) \eta - x \frac{k}{\Omega}\psi + 2 x \eta'
+\frac{k}{\Omega}(6-p)(1-f)(-\chi'+\psi')\right] =0~, \end{array}
& \\[3mm] & \ds
\label{E2} -(6-p)(1-f)x \frac{\Omega}{k}(\chi+\psi)+ 2f\left[
x\frac{\Omega}{k}((8-p)\chi+\psi+x(\chi'+\psi'))+\eta x^2\right]=0
~,~~
& \\[3mm] & \ds
\label{E3} \begin{array}{l} \ds
x^2\bigg(\frac{\Omega^2}{k^2}+f\bigg)\chi-(6-p)(1-f)\frac{\Omega}{k}x\eta
\\[4mm] \ds
+f\left[ -2 x^2 \frac{\Omega}{k} \eta'+(6-p)(1-f)x(\chi'-\psi')
+xf\big((9-p)\chi'+2\psi'+x\chi''\big)\right]=0~, \end{array}
& \\[3mm] & \ds
\label{E4} \begin{array}{l} f\bigg[
\psi\big(-x^2+2(6-p)(1-f)\big)+\chi\big(-x^2+2(8-p)(6-p)(1-f)\big)
\\ \ds
 -2(7-p)x\frac{\Omega}{k}\eta +(6-p)(1-f)x(\chi'+\psi')\bigg]
+f^2\bigg[2(8-p)(6-p)\chi
\\ \ds
+2(6-p)\psi+x\big(3(7-p)\chi'+(7-p)\psi'+x(\chi''+\psi'')\big)\bigg]
-x^2\frac{\Omega^2}{k^2}(\chi+\psi)=0 ~. \end{array} &
\end{eqnarray}
Combining the gauge conditions \eqref{traceeq}-\eqref{transeqs}
with the Einstein equations \eqref{E1}-\eqref{E4} one can derive
Eq.~\eqref{psieq} which is a second order differential equation for $\psi$
of the form
\begin{equation}
\label{E5}
\psi ''(x)+{\cal Q}_p(x) \psi'(x)+{\cal P}_p(x)\psi(x)=0
~.
\end{equation}
${\cal Q}_p$ and ${\cal P}_p$ are $p$-dependent rational functions
of $x$, which we summarize here for completeness
\begin{eqnarray}
{\cal Q}_p(x)&=&f^{-1} x^{-1}
\Big[f^3 k^4 (372-64p+3p^2)+\Omega^2\big(k^2(p-6)^2-4\Omega^2\big)
\nonumber\\
& &+k^2 f^2 \big(k^2(44+4p-2p^2)+3(p-14)(3p-22)\Omega^2\big)
\nonumber\\
& &-k^2 f \big(k^2(p-6)^2+2(292-78p+5p^2)\Omega^2\big)\Big]^{-1}
\nonumber\\
& & \times \Big[k^4 f^4
(-108+124p-21p^2+p^3)-3(p-6)\Omega^2\big(k^2(p-6)^2-4\Omega^2\big)
\nonumber\\
& &+k^2 f^2 \big(k^2(p-6)(-62+5p+p^2)+\Omega^2(11152-4680p+656p^2-31p^3)\big)
\nonumber\\
& &+k^2 f^3 \big(k^2 (2828-964p+106p^2-4p^3)+\Omega^2(-2412+1120p-165p^2+8p^3)\big)
\nonumber\\
& &f \big(2k^4(p-6)^3+k^2 \Omega^2(p-6)(1126-341p+26p^2)+4(11-2p)\Omega^4\big)\Big]
~,
\end{eqnarray}
\begin{eqnarray}
{\cal P}_p(x)&=&k^{-2} f^{-2} x^{-2}
\Big[f^3 k^4 (372-64p+3p^2)+\Omega^2\big(k^2(p-6)^2-4\Omega^2\big)
\nonumber\\
& &+k^2 f^2 \big(k^2(44+4p-2p^2)+3(p-14)(3p-22)\Omega^2\big)
\nonumber\\
& &-k^2 f \big(k^2(p-6)^2+2(292-78p+5p^2)\Omega^2\big)\Big]^{-1}
\nonumber\\
& & \times \Big[-2 k^6 f^5(p-8)^2(p-6)^2+ k^4 f^4 \big( k^2
(8880-110p^3+4p^4-372x^2
\nonumber\\
& &-3p^2(-378+x^2)+p(-5188+64x^2))-(p-6)^2(p-8)(-26+7p)\Omega^2\big)
\nonumber\\
& &-2k^4 f^3\big( k^2(1956-26 p^3+p^4+22x^2+2p(-577+x^2)-p^2(-258+x^2))
\nonumber\\
& &+(-8448+206p^3-9p^4+708x^2+p(6340-144x^2)+p^2(-1734+7x^2))\Omega^2\big)
\nonumber\\
& &+2 fk^2 \Omega^2\big(k^2(p-6)^2(-10-4p+p^2+x^2)+(-84+606x^2+p(26-178x^2)
\nonumber\\
& &+p^2(-2+13x^2))\Omega^2\big)+\Omega^2\big(k^4(p-6)^4-k^2(p-6)^2(-2+7x^2)\Omega^2+4x^2\Omega^4\big)
\nonumber\\
& &+k^2
f^2\big(k^4(p-6)^2(-10+2p+x^2)-2k^2(4992-151p^3+7p^4-294x^2 \nn
\\
\nonumber & &
-6p^2(-199+x^2)+p(-4072+84x^2))\Omega^2+(96-1332x^2+p^2(2-19x^2)
\nonumber\\
& &+4p(-7+80x^2))\Omega^4\big)\Big] ~.
\end{eqnarray}


\section{An off-shell two-parameter family of black branes \label{app:fam}}

In Section \ref{sec:csc}, we described a generalization of the
proof of the correlated stability conjecture by adapting the
analysis of \cite{Reall:2001ag} to the case of (smeared) black
D$p$-branes in the grand canonical ensemble. Part of the argument
involved the construction of an (off-shell) two-parameter family
of Euclidean black branes. In this appendix we consider this
construction in detail.

According to the discussion in Section \ref{sec:csc} we need to
construct a family of black-brane backgrounds with the property
(\ref{aaf}). The backgrounds of this family do not, in general,
satisfy the equations of motion of general relativity, but they
should satisfy the Hamiltonian constraint equations. This is a
special subset of the full equations of motion. For pure gravity
these constraints can be found, for example, in Eqs.\ (10.2.28),
(10.2.30) of \cite{Wald:1984rg}. Our case involves a system of
gravity coupled to a dilaton scalar field and a $(p+2)$-form gauge
field strength. In the Einstein frame the action for this system
is
\begin{equation}
\label{appbaaa} {\cal S}= \frac{1}{16 \pi G} \int d^{10} x
\sqrt{g} \bigg[ -R+\frac{1}{2}(\partial \phi )^2+
\frac{1}{2(p+2)!}e^{\alpha\phi}F^2_{(p+2)}\bigg] ~,
\end{equation}
where $\alpha=\frac{3-p}{2}$. There are two Hamiltonian
constraints for this system.\footnote{There are also ten momentum
constraints, which will be trivially satisfied by the
time-independent family that will be considered in a moment. We
will not discuss them in detail here.} They are
\begin{equation}
\label{appbaab} R_{00}-\frac{1}{2}R g_{00}= \frac{1}{2}(\partial_0
\phi)^2-\frac{1}{4} g_{00} (\partial \phi)^2+
\frac{1}{2(p+1)!}e^{\alpha \phi}\bigg(F_{0 \rho_1 ... \rho_{p+1}}
F_0^{\rho_1 ... \rho_{p+1}}-\frac{1}{2(p+2)}F^2 g_{00}\bigg) ~,
\end{equation}
\begin{equation}
\label{appbaac} \nabla_{\mu}\bigg(e^{\alpha \phi}F^{\mu \rho_1 ...
\rho_p 0}\bigg)=0 ~.
\end{equation}
The first constraint sets the total Hamiltonian to zero and the
second is the Gauss constraint for the gauge field.

Assuming spherical symmetry let us write the generic point in the
family as a background of the form
\begin{equation}
\label{appbaad}
\begin{array}{c} \ds
ds^2=U(r)d\tau^2+V^{-1}(r)dr^2+S(r)\bigg(\sum_{i=1} dx_i^2
+dz^2\bigg)+ R(r) r^2 d\Omega^2_{7-p} ~,
\\[2mm] \ds
 e^{2\phi}=H(r)^{\frac{3-p}{2}}~, ~ ~
F_{(p+2)}=F(r) dt \wedge dr \wedge d x_1 \wedge \cdots \wedge dx_p
~. \end{array}
\end{equation}
The functions $U, V, S, R, H$ and $F$ should be such that the
constraints (\ref{appbaab}), (\ref{appbaac}) are satisfied.
Moreover, the family should include the on-shell smeared
D$p$-branes \eqref{metD0a} (transformed to the Einstein frame),
for which
\begin{equation}
\label{appbaaf} \begin{array}{c} \ds U(r)=H(r)^{-\frac{7-p}{8}}
f(r)~, ~ ~ V(r)=H(r)^{-\frac{p+1}{8}}f(r) \spa
S(r)=R(r)=H(r)^{\frac{p+1}{8}}~,
\\[2mm] \ds
 F(r)=\coth \alpha
\frac{d(H(r)^{-1})}{dr} \spa f(r)=1-\frac{r_0^{6-p}}{r^{6-p}}~, ~
~ H(r)=1+\sinh \alpha \frac{r_0^{6-p}}{r^{6-p}} ~,
\end{array}
\end{equation}
in the Einstein frame.

It is straightforward to plug the ansatz \eqref{appbaad} into the
Hamiltonian constraints (\ref{appbaab}), (\ref{appbaac}) to get a
pair of non-linear (second order and first order) differential
equations for the functions $U, V, S, R, H$ and $F$. These
complicated equations can be simplified considerably with the
following ansatz
\begin{equation}
\label{appbaag} S(r)=R(r)~, ~ ~ V(r)=-H(r)^{\frac{3-p}{4}} U(r) ~,
\end{equation}
which is naturally suggested by the on-shell background
(\ref{appbaaf}). Then, one can solve easily the first order
differential equation that comes from the Gauss constraint
(\ref{appbaac}) to obtain the function $F(r)$ as a functional of
the still arbitrary functions $H(r)$ and $S(r)$. The resulting
expression is
\begin{equation}
\label{appbaagn} F(r)=a H(r)^{\frac{p-3}{2}} S(r)^{p-4} r^{p-7} ~.
\end{equation}
Here $a$ is an integration constant that will be fixed
appropriately below. Plugging this expression back into the
differential equation that originates from (\ref{appbaab}) gives
the differential equation
\begin{equation}
\label{appbaah} U'(r)+P[H,S](r)U(r)+Q[H,S](r)=0 ~,
\end{equation}
where $'$ denotes differentiation with respect to $r$ and
$P[H,S]$, $Q[H,S]$ are the following functionals of the still
arbitrary functions $H$ and $S$
\begin{equation}
\label{appbaai}
\begin{array}{rcl}
P[H,S]&=&  \bigg[-(\ln \Lambda)''-\frac{1}{4} \big( (\ln
\Lambda)'\big)^2+ \frac{3-p}{4}(\ln H)''-\frac{(3-p)^2}{64}\big(
(\ln H)' \big)^2
\\
&&  -\frac{p+1}{4}\big( (\ln S)'\big)^2-\frac{7-p}{4}\big( (\ln
(Sr^2))'\big)^2\bigg] \bigg[-\frac{1}{2} (\ln
\Lambda)'+\frac{3-p}{8}(\ln H)'\bigg]^{-1} ~,
\\[3mm]
Q[H,S]&=&  -\bigg[(7-p)(6-p)S^{-1}H^{\frac{p-3}{4}}r^{-2}+ a^2
\frac{p+3}{2(p+2)!}H^{\frac{(p-3)(p+5)}{8}}S^{p-10}r^{2(p-7)}
\bigg]\\
&&  \times \bigg[-\frac{1}{2} (\ln \Lambda)'+\frac{3-p}{8}(\ln
H)'\bigg]^{-1} ~. \end{array}
\end{equation}
The functional $\Lambda[H,S]$ is defined by the equation
\begin{equation}
\Lambda[H,S]=H^{\frac{3-p}{4}}S^8 r^{2(7-p)}
~.
\end{equation}
Equation (\ref{appbaah}) can be solved with standard methods
to obtain the function $U(r)$ in terms of $H(r)$ and $S(r)$. The solution is
\begin{equation}
\label{appbaaj} U(r)=-e^{-\int^r P[H,S](t)dt} \bigg\{ \int^r
e^{\int^s P[H,S](t)dt} Q[H,S](s)ds +b\bigg\} ~,
\end{equation}
where $b$ is again an integration constant.

This analysis shows that it is possible to construct an off-shell
two-parameter family of smeared black D$p$-branes that are
spherically symmetric and respect the Hamiltonian constraints. The
family is expressed in terms of two free functions $H(r)$ and
$S(r)$ (compare this to the magnetically charged unsmeared case of
\cite{Reall:2001ag}, where one considers a one-parameter family
expressed in terms of one free function). The only restrictions
that should be applied to the functions $H$ and $S$ are certain
boundary conditions at the horizon and boundary at infinity, which
are explained in the main text. The integration constants $a$ and
$b$ above are fixed by these boundary conditions.

\end{appendix}

\addcontentsline{toc}{section}{References}

\providecommand{\href}[2]{#2}\begingroup\raggedright\endgroup


\begin{thebibliography}{10}

\bibitem{Aharony:1999ti}
O.~Aharony, S.~S. Gubser, J.~Maldacena, H.~Ooguri, and Y.~Oz,
``Large {$N$}
  field theories, string theory and gravity,'' {\em Phys. Rept.} {\bf 323}
  (2000) 183,
\href{http://www.arXiv.org/abs/hep-th/9905111}{{\tt
hep-th/9905111}}.

\bibitem{Gregory:1993vy}
R.~Gregory and R.~Laflamme, ``Black strings and {$p$}-branes are
unstable,''
  {\em Phys. Rev. Lett.} {\bf 70} (1993) 2837--2840,
\href{http://arXiv.org/abs/hep-th/9301052}{{\tt hep-th/9301052}}.

\bibitem{Gregory:1994bj}
R.~Gregory and R.~Laflamme, ``The instability of charged black
strings and
  p-branes,'' {\em Nucl. Phys.} {\bf B428} (1994) 399--434,
\href{http://arXiv.org/abs/hep-th/9404071}{{\tt hep-th/9404071}}.

\bibitem{Horowitz:2001cz}
G.~T. Horowitz and K.~Maeda, ``Fate of the black string
instability,'' {\em
  Phys. Rev. Lett.} {\bf 87} (2001) 131301,
\href{http://arXiv.org/abs/hep-th/0105111}{{\tt hep-th/0105111}}.

\bibitem{Harmark:2005pp}
T.~Harmark and N.~A. Obers, ``Phases of {Kaluza-Klein} black
holes: {A} brief
  review,''
\href{http://www.arXiv.org/abs/hep-th/0503020}{{\tt
hep-th/0503020}}.

\bibitem{Kol:2004ww}
B.~Kol, ``The phase transition between caged black holes and black
strings: {A}
  review,''
\href{http://www.arXiv.org/abs/hep-th/0411240}{{\tt
hep-th/0411240}}.

\bibitem{Gubser:2000ec}
S.~S. Gubser and I.~Mitra, ``Instability of charged black holes in
anti-{de
  Sitter} space,''
\href{http://www.arXiv.org/abs/hep-th/0009126}{{\tt
hep-th/0009126}}.

\bibitem{Gubser:2000mm}
S.~S. Gubser and I.~Mitra, ``The evolution of unstable black holes
in anti-de
  {Sitter} space,'' {\em JHEP} {\bf 08} (2001) 018,
\href{http://arXiv.org/abs/hep-th/0011127}{{\tt hep-th/0011127}}.

\bibitem{Reall:2001ag}
H.~S. Reall, ``Classical and thermodynamic stability of black
branes,'' {\em
  Phys. Rev.} {\bf D64} (2001) 044005,
\href{http://arXiv.org/abs/hep-th/0104071}{{\tt hep-th/0104071}}.

\bibitem{Gregory:2001bd}
J.~P. Gregory and S.~F. Ross, ``Stability and the negative mode
for
  {Schwarzschild} in a finite cavity,'' {\em Phys. Rev.} {\bf D64} (2001)
  124006,
\href{http://www.arXiv.org/abs/hep-th/0106220}{{\tt
hep-th/0106220}}.

\bibitem{Gregory:1994tw}
R.~Gregory and R.~Laflamme, ``Evidence for stability of extremal
black
  {$p$}-branes,'' {\em Phys. Rev.} {\bf D51} (1995) 305--309,
\href{http://www.arXiv.org/abs/hep-th/9410050}{{\tt
hep-th/9410050}}.

\bibitem{Hirayama:2002hn}
T.~Hirayama, G.-w. Kang, and Y.-o. Lee, ``Classical stability of
charged black
  branes and the {Gubser-Mitra} conjecture,'' {\em Phys. Rev.} {\bf D67} (2003)
  024007,
\href{http://www.arXiv.org/abs/hep-th/0209181}{{\tt
hep-th/0209181}}.

\bibitem{Kang:2004hm}
G.~Kang and J.~Lee, ``Classical stability of black {D3}-branes,''
{\em JHEP}
  {\bf 03} (2004) 039,
\href{http://www.arXiv.org/abs/hep-th/0401225}{{\tt
hep-th/0401225}}.

\bibitem{Kang:2004ys}
G.~Kang, ``Classical stability of black branes,'' {\em J. Korean
Phys. Soc.}
  {\bf 45} (2004) S86--S89,
\href{http://www.arXiv.org/abs/hep-th/0403015}{{\tt
hep-th/0403015}}.

\bibitem{Gubser:2004dr}
S.~S. Gubser, ``The {Gregory-Laflamme} instability for the {D2-D0}
bound
  state,'' {\em JHEP} {\bf 02} (2005) 040,
\href{http://www.arXiv.org/abs/hep-th/0411257}{{\tt
hep-th/0411257}}.

\bibitem{Ross:2005vh}
S.~F. Ross and T.~Wiseman, ``Smeared {D0} charge and the
{Gubser-Mitra}
  conjecture,'' {\em Class. Quant. Grav.} {\bf 22} (2005) 2933--2946,
\href{http://www.arXiv.org/abs/hep-th/0503152}{{\tt
hep-th/0503152}}.

\bibitem{Friess:2005tz}
J.~J. Friess and S.~S. Gubser, ``Instabilities of {D-brane} bound
states and
  their related theories,''
\href{http://www.arXiv.org/abs/hep-th/0503193}{{\tt
hep-th/0503193}}.

\bibitem{Friess:2005zp}
J.~J. Friess, S.~S. Gubser, and I.~Mitra, ``Counter-examples to
the correlated
  stability conjecture,''
\href{http://www.arXiv.org/abs/hep-th/0508220}{{\tt
hep-th/0508220}}.

\bibitem{Harmark:2002tr}
T.~Harmark and N.~A. Obers, ``Black holes on cylinders,'' {\em
JHEP} {\bf 05}
  (2002) 032,
\href{http://www.arXiv.org/abs/hep-th/0204047}{{\tt
hep-th/0204047}}.

\bibitem{Harmark:2004ws}
T.~Harmark and N.~A. Obers, ``New phases of near-extremal branes
on a circle,''
  {\em JHEP} {\bf 09} (2004) 022,
\href{http://www.arXiv.org/abs/hep-th/0407094}{{\tt
hep-th/0407094}}.

\bibitem{Bostock:2004mg}
P.~Bostock and S.~F. Ross, ``Smeared branes and the {Gubser-Mitra}
  conjecture,'' {\em Phys. Rev.} {\bf D70} (2004) 064014,
\href{http://www.arXiv.org/abs/hep-th/0405026}{{\tt
hep-th/0405026}}.

\bibitem{Aharony:2004ig}
O.~Aharony, J.~Marsano, S.~Minwalla, and T.~Wiseman, ``Black hole
- black
  string phase transitions in thermal 1+1 dimensional supersymmetric
  {Yang-Mills} theory on a circle,'' {\em Class. Quant. Grav.} {\bf 21} (2004)
  5169--5192,
\href{http://www.arXiv.org/abs/hep-th/0406210}{{\tt
hep-th/0406210}}.

\bibitem{Kudoh:2005hf}
H.~Kudoh and U.~Miyamoto, ``On non-uniform smeared black branes,''
\href{http://www.arXiv.org/abs/hep-th/0506019}{{\tt
hep-th/0506019}}.

\bibitem{Harmark:2005pq}
T.~Harmark and N.~A. Obers, ``New phases of thermal {SYM} and
{LST} from
  {Kaluza-Klein} black holes,'' {\em Fortsch. Phys.} {\bf 53} (2005) 536--541,
\href{http://www.arXiv.org/abs/hep-th/0503021}{{\tt
hep-th/0503021}}.

\bibitem{Gregory:1988nb}
R.~Gregory and R.~Laflamme, ``Hypercylindrical black holes,'' {\em
Phys. Rev.}
  {\bf D37} (1988)
305.

\bibitem{Gubser:2001ac}
S.~S. Gubser, ``On non-uniform black branes,'' {\em Class. Quant.
Grav.} {\bf
  19} (2002) 4825--4844,
\href{http://www.arXiv.org/abs/hep-th/0110193}{{\tt
hep-th/0110193}}.

\bibitem{Wiseman:2002zc}
T.~Wiseman, ``Static axisymmetric vacuum solutions and non-uniform
black
  strings,'' {\em Class. Quant. Grav.} {\bf 20} (2003) 1137--1176,
\href{http://www.arXiv.org/abs/hep-th/0209051}{{\tt
hep-th/0209051}}.

\bibitem{Sorkin:2004qq}
E.~Sorkin, ``A critical dimension in the black-string phase
transition,'' {\em
  Phys. Rev. Lett.} {\bf 93} (2004) 031601,
\href{http://www.arXiv.org/abs/hep-th/0402216}{{\tt
hep-th/0402216}}.

\bibitem{Sarbach:2004rm}
O.~Sarbach and L.~Lehner, ``Critical bubbles and implications for
critical
  black strings,'' {\em Phys. Rev.} {\bf D71} (2005) 026002,
\href{http://www.arXiv.org/abs/hep-th/0407265}{{\tt
hep-th/0407265}}.

\bibitem{Harmark:2004bb}
T.~Harmark and N.~A. Obers. Work in progress.

\bibitem{Kang:2005is}
G.~Kang and Y.~Lee, ``Stability of smeared black branes and the
{Gubser-Mitra}
  conjecture,''
\href{http://www.arXiv.org/abs/hep-th/0504031}{{\tt
hep-th/0504031}}.

\bibitem{Hassan:1992mq}
S.~F. Hassan and A.~Sen, ``Twisting classical solutions in
heterotic string
  theory,'' {\em Nucl. Phys.} {\bf B375} (1992) 103--118,
\href{http://www.arXiv.org/abs/hep-th/9109038}{{\tt
hep-th/9109038}}.

\bibitem{Hawking:1996fd}
S.~W. Hawking and G.~T. Horowitz, ``The gravitational
{Hamiltonian}, action,
  entropy and surface terms,'' {\em Class. Quant. Grav.} {\bf 13} (1996)
  1487--1498,
\href{http://www.arXiv.org/abs/gr-qc/9501014}{{\tt
gr-qc/9501014}}.

\bibitem{Hawking:1995ap}
S.~W. Hawking and S.~F. Ross, ``Duality between electric and
magnetic black
  holes,'' {\em Phys. Rev.} {\bf D52} (1995) 5865--5876,
\href{http://www.arXiv.org/abs/hep-th/9504019}{{\tt
hep-th/9504019}}.

\bibitem{Whiting:1988qr}
B.~F. Whiting and J.~York, James~W., ``Action principle and
partition function
  for the gravitational field in black hole topologies,'' {\em Phys. Rev.
  Lett.} {\bf 61} (1988)
1336.

\bibitem{Prestidge:1999uq}
T.~Prestidge, ``Dynamic and thermodynamic stability and negative
modes in
  {Schwarzschild-anti-de Sitter},'' {\em Phys. Rev.} {\bf D61} (2000) 084002,
\href{http://www.arXiv.org/abs/hep-th/9907163}{{\tt
hep-th/9907163}}.

\bibitem{Buchel:2005nt}
A.~Buchel, ``A holographic perspective on {Gubser-Mitra}
conjecture,'' \href{http://www.arXiv.org/abs/hep-th/0507275}{{\tt
hep-th/0507275}}.

\bibitem{Adams:2005rb}
A.~Adams, X.~Liu, J.~McGreevy, A.~Saltman, and E.~Silverstein,
``Things fall
  apart: Topology change from winding tachyons,''
\href{http://www.arXiv.org/abs/hep-th/0502021}{{\tt
hep-th/0502021}}.

\bibitem{Horowitz:2005vp}
G.~T. Horowitz, ``Tachyon condensation and black strings,''
\href{http://www.arXiv.org/abs/hep-th/0506166}{{\tt
hep-th/0506166}}.

\bibitem{Aharony:2005ew}
O.~Aharony {\em et al.}, ``The phase structure of low dimensional
large {$N$}
  gauge theories on tori,''
\href{http://www.arXiv.org/abs/hep-th/0508077}{{\tt
hep-th/0508077}}.

\bibitem{Gross:1982cv}
D.~J. Gross, M.~J. Perry, and L.~G. Yaffe, ``Instability of flat
space at
  finite temperature,'' {\em Phys. Rev.} {\bf D25} (1982)
330--355.

\bibitem{Kol:2004pn}
B.~Kol and E.~Sorkin, ``On black-brane instability in an arbitrary
dimension,''
  {\em Class. Quant. Grav.} {\bf 21} (2004) 4793--4804,
\href{http://www.arXiv.org/abs/gr-qc/0407058}{{\tt
gr-qc/0407058}}.

\bibitem{Wald:1984rg}
R.~M. Wald, {\em General Relativity}.
\newblock The University of Chicago Press, 1984.

\end{thebibliography}

\end{document}